\documentclass[%
reprint,
superscriptaddress,
amsmath,amssymb,
]{revtex4-1}

\usepackage{graphicx}
\usepackage{dcolumn}
\usepackage{bm}
\usepackage{hyperref}

\newcommand{\calH}{\mathcal{H}}

\newcommand{\eff}{\mathrm{eff}}

\usepackage{braket}
\usepackage{color}
\usepackage{ulem}

\begin{document}

\preprint{APS/123-QED}

\title{Control of Magnetic and Topological Orders with a DC Electric Field}
\author{Kazuaki Takasan}
\email{takasan@scphys.kyoto-u.ac.jp}
\affiliation{%
 Department of Physics, Kyoto University, Kyoto 606-8502, Japan
}%
\author{Masahiro Sato}
\email{masahiro.sato.phys@vc.ibaraki.ac.jp}
\affiliation{%
Department of Physics, Ibaraki University, Mito, Ibaraki 310-8512, Japan
}%

\date{\today}

\begin{abstract}
We theoretically propose a new route to control magnetic and topological orders in a broad class of insulating magnets 
with a DC electric field. We show from the strong-coupling expansion 
that magnetic exchange interactions along the electric-field direction are generally enhanced in Mott insulators.  
We demonstrate that several magnetic or topological ordered phases such as quantum spin liquids 
and Haldane-gap states can be derived if we apply a strong enough DC electric field to typical frustrated or low-dimensional magnets. 
Our proposal is effective especially for weak Mott insulators and magnets in the vicinity of quantum critical points, 
and would also be applicable for magnets under low-frequency AC electric fields such as terahertz laser pulses. 
A similar strategy of controlling exchange interactions can also be utilized in cold atomic systems.
\end{abstract}

\maketitle

\textit{Introduction. ---}
One of the most important goals in condensed matter physics is to control the quantum states of matter. 
In recent years, great efforts have been made to understand how to control solid states by external fields 
both theoretically and experimentally. Particularly, many scenarios with AC electromagnetic fields or laser light, 
including the control of topological \cite{Oka2009, Lindner2011, Grushin2014, Sato2014, Wang2013, Jotzu2014}, 
magnetic \cite{Takayoshi2014a, Takayoshi2014b, Mentink2015, Sato2016, Kitamura2017, Claassen2017, Eckstein2017, Gorg2018}, 
and superconducting \cite{Knap2016, Murakami2017, Takasan2017, Fausti2011, Mitrano2016, Matsunaga2014} orders, have been proposed 
and gathering much attention. For instance, some signatures of the realization of AC-field driven topological insulators 
(called Floquet topological insulators) have been detected in recent years \cite{Wang2013, Jotzu2014}. 
The control with low-frequency or DC (static) electromagnetic fields 
has been also studied intensively. 
For example, electric-field-controlled magnetism in multiferroics~\cite{Kimura2003, Katsura2005, Mostovoy2006, Pimenov2006, Tokunaga2012, Tokura2014}
 and Mott breakdown driven by DC electric fields~\cite{Fukui1998, Oka2003, Eckstein2010, Oka2012, Yamakawa2017, Sow2017} 
 are two of the attractive topics in the research field of DC-field control.
 
However, a wider range of DC-field driven phenomena has not been explored well compared with AC-field studies.  
DC fields usually do not make the system heated, while it is difficult to avoid heating effect in AC-field driven systems. 
This is a significant advantage of the DC-field study. 
Moreover, in recent years, experimental ways of generating strong DC electric fields (e.g., order of 1-10~MV/cm) have been developed by using several techniques based on, for example, field-effect transistors~\cite{Ueno2014, Bisri2017} and nano-scale needles~\cite{Hsu2016}. 
The technology of low-frequency AC fields has also been developed and 
for instance we can use terahertz (THz) laser pulses whose intensity is the order of 1-10~MV/cm~\cite{Hirori2011, Nicoletti2016}.
Novel proposals for DC-field and low-frequency AC-field driven phenomena thereby are being anticipated.  

\begin{figure}[t]
\includegraphics[width=8.4cm]{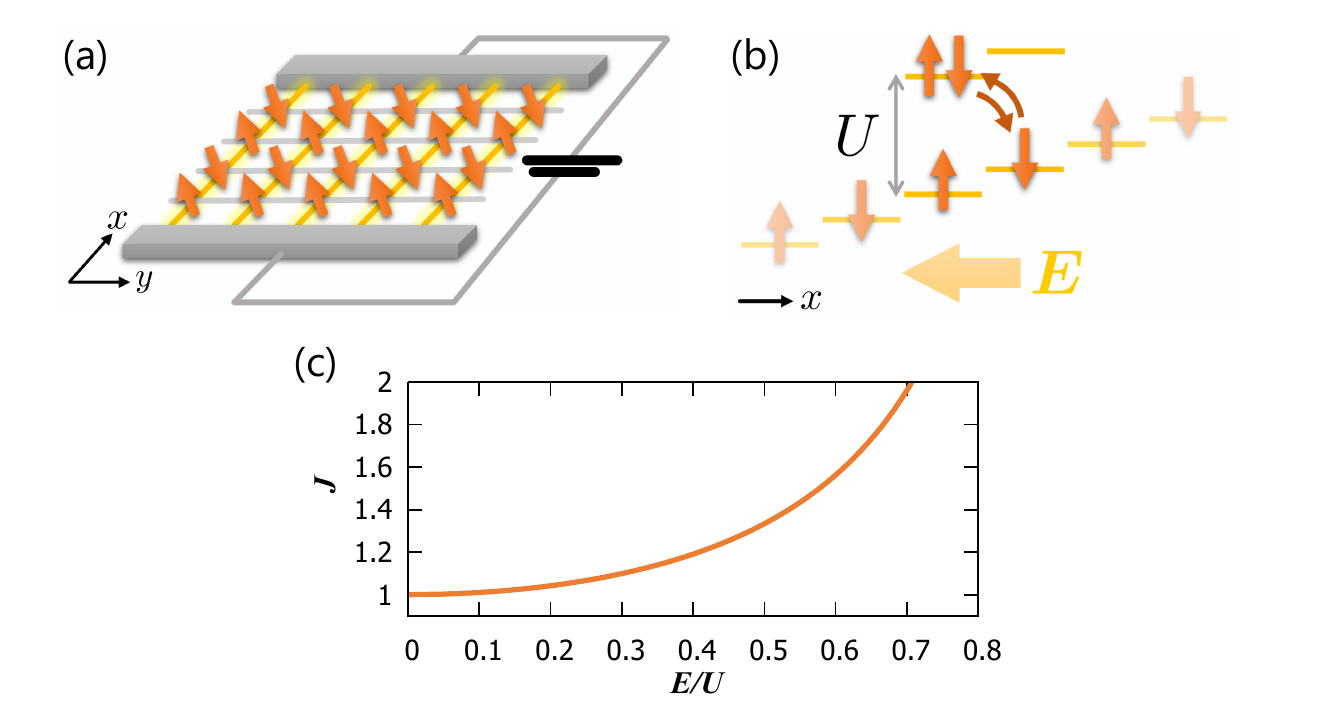}
\caption{
(a) Our setup of a Mott insulator under DC electric fields $\bm E$ along the $x$-direction. 
(b) Spatial energy level structure of the DC-field driven Mott insulator.  Arrows denote the second-order virtual hopping 
process in the Mott state. (c) Electric-field dependence of the exchange coupling along the $x$-direction 
in the Mott system (b) [see Eq.~(\ref{eq:spin-1/2})].
}
\label{setup}
\end{figure}

In this Letter, we theoretically show that low-frequency or DC electric fields have a high potential 
to generate rich magnetic states in solids. 
We propose a new way to control magnetic or topological orders in Mott insulators by static electric fields. 
We consider quantum magnets originating from Mott insulators with DC electric fields as shown in Fig. \ref{setup} (a). 
In this setup, we derive the low-energy effective spin models by applying the strong-coupling expansion, 
and show that exchange interactions along the DC-field direction are generally increased with the growth of the field strength. 
A strong electric field comparable to Mott gap is usually necessary for realizing Mott breakdown, 
while our proposal indicates that quantum magnetic nature can be changed with smaller DC fields in Mott insulators. 
We show that various quantum states such as quantum spin liquids \cite{Savary2017, Zhou2017, XGWen_book} 
 and Haldane-gap states \cite{Haldane1983a, Haldane1983b, Affleck1987, Affleck_Lecture, Pollmann2010, Pollmann2012, Giamarchi_book, Gogolin_book, Tsvelik_book} 
 can be created/annihilated by applying strong DC electric fields 
to representative frustrated or quasi-one-dimensional (quasi-1D) magnets.  

\textit{Enhancement of the Exchange Coupling. ---}
To show how the exchange interaction is modified by DC electric fields, we first consider 
a generic half-filled, single-band Hubbard model subject to static electric fields $\bm E$. 
The effect of electric fields is introduced as an on-site potential and the Hamiltonian is given by
\begin{align}
\calH &= \sum_{\bm r \bm r^\prime \sigma}  t_{\bm r \bm r^\prime} c^\dagger_{\bm r  \sigma} c_{\bm r^\prime \sigma}  
+ U \sum_{\bm r} n_{\bm r \uparrow} n_{\bm r \downarrow} 
+ \sum_{\bm r \sigma} V_{\bm r}  n_{\bm r \sigma}, 
\label{Model_General}
\end{align}
where $c_{\bm r\sigma}$ is a spin-$\sigma$ electron annihilation operator ($\sigma=\uparrow,\downarrow$) 
on a site $\bm r=(i,j,k)$, and $n_{\bm r \sigma}=c^\dag_{\bm r  \sigma}c_{\bm r \sigma}$ 
(the lattice constant is set to be unity).  
The first and second terms respectively stand for hopping and on-site Coulomb repulsion, and 
the on-site potential $V_{\bm r}$ represents the effect of the applied electric field. 
For example, $V_{\bm r}$ is reduced to $V_i = i |\bm E|=i E$ when the electric field is parallel to the $x$-axis
~\cite{FN1}.
If the Coulomb repulsion $U>0$ is strong enough, a Mott insulator is realized and 
we can derive its low-energy effective spin model by treating the kinetic term as a perturbation (large $U$ expansion). 
The point is that the second-order virtual hopping amplitude becomes direction-dependent 
due to the field-driven potential $V_{\bm r}$  as shown in Fig.~\ref{setup}~(b). 
As a result, the exchange interaction becomes spatially anisotropic and the effective Hamiltonian 
in the second-order perturbation is given by  
\begin{align}
\calH_\eff = \sum_{\langle \bm r, \bm r^\prime \rangle} 
\frac{J_{\bm r \bm r^\prime}}{1- \left( \frac{\Delta V_{\bm r \bm r^\prime}}{U} \right)^2} 
\bm S_{\bm r} \cdot \bm S_{\bm r^\prime},
\label{eq:spin-1/2}
\end{align}
where $\bm S_{\bm r}$ is the electron spin operator on a site $\bm r$, 
$J_{\bm r \bm r^\prime}=4 |t_{\bm r \bm r^\prime}|^2/U$ and $\Delta V_{\bm r \bm r^\prime} = V_{\bm r}-V_{\bm r^\prime}$. The summation is taken over all the bonds $\langle \bm r, \bm r^\prime \rangle$.
The perturbation expansion would be valid if the on-site potential energy is smaller than the Mott gap, i.e., $|V_{\bm r}| \alt U$
~\cite{FN2}.
This effective spin-$\frac{1}2$ Heisenberg model clearly shows that antiferromagnetic (AFM) exchange couplings are 
generally enhanced by the DC electric field. 
For example, if we focus on a bond $\langle \bm r_1, \bm r_2 \rangle$ parallel 
to the electric-field direction, the potential difference $\Delta V_{\bm r_1 \bm r_2}$ 
is given by $E |\bm r_1-\bm r_2|$ and thereby the exchange coupling on the nearest-neighboring bond is 
computed as $J/(1-(E/U)^2)$ with $J=4t^2/U$ being the exchange coupling in the original Hubbard model 
without any potential $V_{\bm r}$ [Fig.~\ref{setup} (c)].

The above argument and the physical picture in Fig.~\ref{setup}~(b) clearly indicates that the DC-field driven enhancement 
of exchange couplings generally takes place in a quite wide class of Mott insulators~\cite{FN3}.
The scope is not limited to solid state systems. 
Our results are also applicable to Mott insulators in ultracold atoms on optical lattices~\cite{Bloch2008}. 
Tilting optical-lattice potentials plays the same role as the DC field in solid systems~\cite{Greiner2002}, 
and it is relatively easy to create such a tilted optical lattice. 
The tilt would be useful for realizing/controlling an AFM long range order in cold atoms~\cite{Bloch2008, Mazurenko2017}.

If a similar strategy of the perturbation theory is applied to a half-filled two-orbital Hubbard model, 
we obtain the following spin-1 AFM Heisenberg model
\begin{align}
\calH_\eff =\sum_{\langle \bm r, \bm r^\prime \rangle} 
\frac{ J^\prime_{\bm r \bm r^\prime}}{1 - \left(\frac{\Delta V_{\bm r \bm r^\prime}}{U+J_H}\right)^2} 
\bm S_{\bm r} \cdot \bm S_{\bm r^\prime},
\label{eq:spin-1}
\end{align}
where $\bm S_{\bm r}$ is the spin-1 operator on a site $\bm r$,  
$J^\prime_{\bm r \bm r^\prime}=2 |t_{\bm r \bm r^\prime}|^2/(U+J_H)$ and 
$J_H$ is the strength of the Hund's coupling~\cite{FN4}.
We stress that both the effective models (\ref{eq:spin-1/2}) and (\ref{eq:spin-1}) could be relevant 
even in a short period when a sufficiently low frequency AC electric field (e.g., THz laser pulse) 
is applied to the Mott insulators 
instead of DC fields~\cite{FN5}.
We also note that on top of exchange couplings, spin-orbit (SO) couplings can be changed by DC electric fields 
\cite{Nitta1997, Caviglia2010, Soumyanarayanan2016}, although their strength would strongly depend on the detail of atomic wave functions 
and lattice structures.

\begin{figure}[t]
\includegraphics[width=8cm]{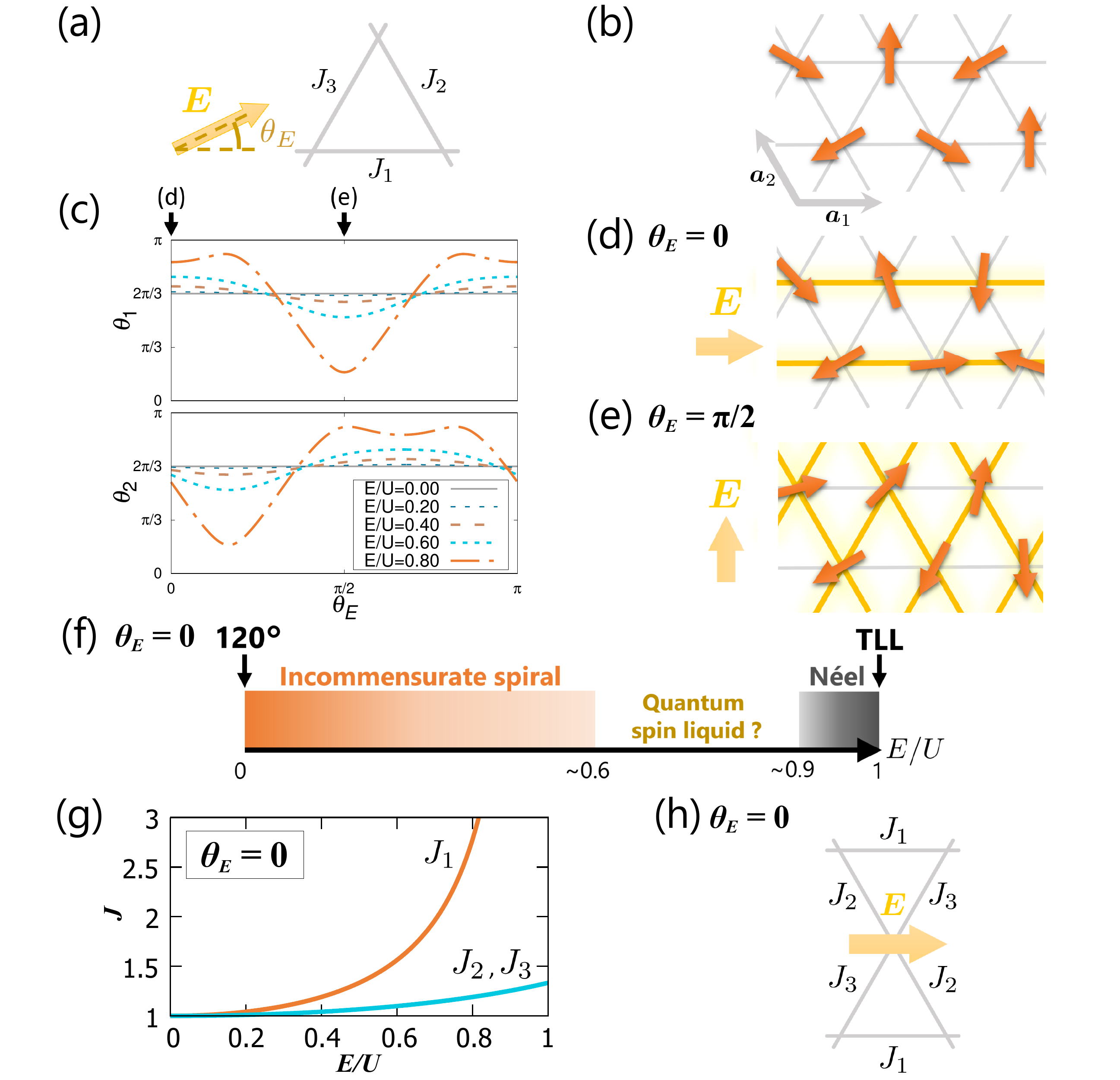}
\caption{(a) Exchange couplings $J_{1, 2, 3}$ in a triangular lattice and the angle $\theta_E$ of the DC electric field $\bm E$. 
(b) 120$^\circ$ structure, which is the ground state of the triangular AFM Heisnberg model. 
(c) $\bm E$ dependence of pitch angles $\theta_1$ and $\theta_2$ in the spiral ordered phase 
of the triangular AFM model (\ref{eq:triangular}) in DC electric fields. 
(d, e) Typical spin configurations in the $\bm E$-driven spiral ordered phases for (d) $\theta_E=0$ and (e) $\pi/2$. 
(f) Ground-state phase diagram of the triangular spin-$\frac{1}{2}$ AFM Heisenberg model 
in DC electric fields ($\theta_E=0$). 
(g) $\bm E$ dependence of the exchange couplings $J_{1,2,3}$ in the model~(\ref{eq:triangular}).
(h) Exchange couplings $J_{1,2,3}$ in a Kagom\'{e} lattice magnet with DC electric fields $\bm E$.}
\label{tri}
\end{figure}

\textit{DC-field Driven Phases and Transitions ---}
On the basis of the above perturbation theory, we show how magnetic properties of Mott insulators can be controlled 
by DC electric fields. 
For instance, weak Mott insulators and magnets residing around critical points are expected to be quite relevant 
for the purpose of the DC-field control since their quantum states are unstable against a small change of magnetic interactions.  
\textit{Frustrated Magnets. ---}
In frustrated magnets, spatial structures of magnetic interactions 
determine their magnetic orders, and the modification of the spatial structures with DC electric fields 
enables us to change the orders. Namely, frustrated magnets are expected to give one of the best stages
for electric-field control of magnetism.

First, we consider a spin-$\frac{1}2$ AFM Heisenberg model on a triangular lattice as a typical frustrated magnet. 
If we apply DC fields to a spatially-isotropic Mott insulating triangular magnet, 
the spin Hamiltonian is given as 
\begin{align}
\calH_\mathrm{tri} &= \sum_{\bm r} \sum^3_{k=1}  J_k (E, \theta_E)   \bm S_{\bm r} \cdot \bm S_{\bm r + \bm a_k}.
\label{eq:triangular}
\end{align}
Here the vector $\bm r$ denotes a site on the triangular lattice, and primitive translation vectors $\bm a_{1,2,3}$ 
are given by $\bm a_1=(1,0)$, $\bm a_2=(-1/2,\sqrt{3}/2)$ and $\bm a_3=\bm a_1+\bm a_2$ [Fig.~\ref{tri} (b)]. 
The direction of the applied DC field $\bm E$ is controlled with the angle $\theta_E$ as in Fig.~\ref{tri} (a).  
The parameter $J_k$ represent the strength of the exchange coupling parallel to $\bm a_k$ ($k=1, 2, 3$) 
and their $E$ dependence is computed as $J_1(E,\theta_E)=J/[1-\{E\cos\theta_E/U\}^2]$, 
$J_2 (E,\theta_E)=J/[1- \{E\cos (\theta_E-2\pi/3) / U\}^2]$, 
and $J_3(E,\theta_E)=J/[1-\{E\cos (\theta_E-\pi/3)/U\}^2]$ with $J=4t^2/U$.

Without electric fields ($E=0$), the ground state of this model is 
a commensurate 120$^\circ$ structure shown in Fig.~\ref{tri} (b) \cite{Jolicoeur1989, Neuberger1989, Bernu1994}. 
When a field $\bm E$ is applied, the exchange coupling becomes anisotropic, 
and an incommensurate spiral order would emerge. 
From the simple calculation of the classical ground state energy, we can determine the pitch angle of 
the incommensurate state as a function of $E(=|\bm E|)$ and $\theta_E$. 
Figure~\ref{tri} (c) depicts the pitch angle $\theta_{1(2)}$ that is defined as the 
difference between two neighboring spins' angle on the bond along the $\bm a_{1(2)}$ direction. 
As shown in Fig.~\ref{tri} (d) and (e), if $\theta_E$ is locked to zero ($\pi/2$), 
the one dimensionality is enhanced (the system is gradually changed into a square lattice system). 
These results clearly indicate that the spiral order pattern can be controlled by electric fields smaller than 
the critical value of the Mott breakdown. 

If we focus on the case of $\theta_E=0$, the system is a spin-$\frac{1}{2}$ anisotropic triangular lattice model with $J_2=J_3$, 
and it has been well studied both theoretically and experimentally \cite{Zhou2017}. 
Some previous studies \cite{Weihong1999, Yunoki2006, Weng2006, Heidarian2009, Harada2012} show 
that the spiral order is preserved at least up to $J_2/J_1\sim 0.6$ 
when $J_1$ is increased with $E$. On the other hand, a reliable approach based on 1D quantum field theory 
shows that a N\'{e}el order should appear near the anisotropic limit ($J_2 / J_1 \to 0$) \cite{Starykh2007}. 
At the point of $J_2=0$, the system is  reduced to decoupled 1D Heisenberg chains and 
a Tomonaga-Luttinger liquid (TLL) phase appears. 
The quantum phases between spiral and N\'{e}el orders are still under debate, 
but it is predicted to be a quantum spin liquid \cite{Yunoki2006, Weng2006, Heidarian2009, Harada2012}. 
Combining these results with the $E$ dependence of $J_2/J_1$, 
we obtain the ground-state phase diagram under the electric field $\bm E$ with $\theta_E=0$, 
as shown in Fig. \ref{tri} (f). 
Note that the end point of the N\'{e}el order has never been theoretically determined.

Our approach indicates that sufficiently strong electric fields $E/U \sim 0.6$ are necessary for the emergence of 
quantum spin liquid states if we start from the isotropic point $J_{1,2,3}=J$ at $\bm E=\bm 0$. 
This critical strength of the electric field corresponds to $\sim$ 5 MV/cm 
for typical organic triangular Mott insulators, e.g. (ET)$_2$Cu(NCS)$_2$~\cite{Schultz1991, Nakamura2009} 
and (ET)$_2$Cu$_2$(CN)$_3$~\cite{Shimizu2003, Kezsmarki2006, Nakamura2009}, 
and it is in principle possible to reach this value 
by relying current techniques such as field-effect transistors \cite{Mannhart1993, Ueno2014}.

In addition to the triangular lattice system, 
here we give a few remarks on the Kagom\'{e} lattice magnets. 
One sees from Fig.~\ref{tri} (g) that if we apply an electric field to a spatially isotropic Kagom\'{e} Mott insulator, 
three kinds of exchange couplings $J_{1,2,3}$ appear. 
The $\bm E$ dependence of $J_{1,2,3}$ is completely same as that of the triangular lattice.  
Our method provides the way of generating anisotropic Kagom\'{e} lattices. 

\begin{figure}[t]
\includegraphics[width=7.8cm]{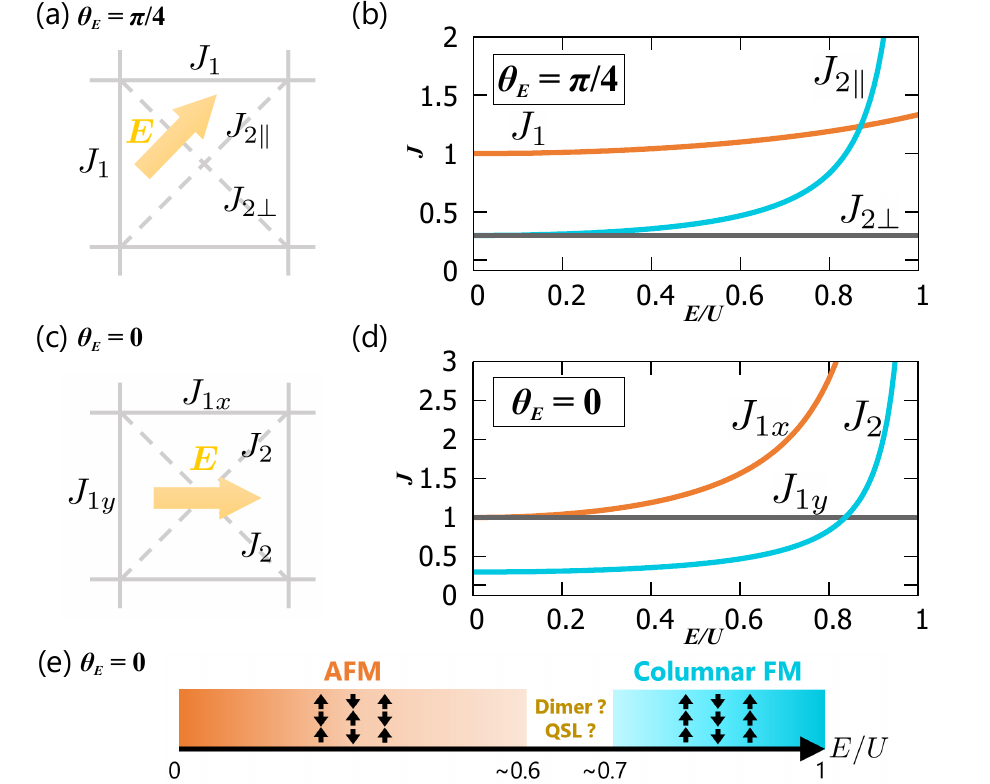}
\caption{(a, c) Exchange couplings in DC-field driven $J_1$-$J_2$ square-lattice magnets 
for (a) $\theta_E=\pi/4$ and (c) $\theta_E=0$.  
In the case of $\theta_E=\pi/4$ ($0$), $J_2$ ($J_1$) is changed into $J_{2\parallel}$ and $J_{2\perp}$ ($J_{1x}$ and $J_{1y}$). 
(b, d) $\bm E$ dependence of the exchange couplings in the case of (b) $\theta_E=\pi/4$ and (d) $\theta_E=0$. 
(e) Ground-state phase diagram of the AFM spin-$\frac{1}2$ Heisenberg model on a $J_1$-$J_2$ square lattice 
in DC electric fields ($\theta_E=0$). In Panels (b), (d), and (e), we set $J_2/J_1=0.3$ at $E = 0$.}
\label{frust}
\end{figure}

Next, we turn to the spin-$\frac{1}{2}$ magnet on a $J_1$-$J_2$ square lattice.
On the top of the triangular magnet, this model is another representative of 2D frustrated systems 
and has been long studied \cite{Starykh2004,Bombardi2005, Nath2008, Tsirlin2008, Oka2008, Jiang2012, Metavitsiadis2014}. 
We calculate how the exchange couplings are modified by an electric field parallel to a $J_1$ ($J_2$) bond, 
as shown in Fig.~\ref{frust} (a)-(d). 
In the case of $\bm E$ parallel to the $J_{1x}$ bond ($\theta_E=0$), one dimensionality is enhanced along the $J_{1x}$ bond direction, 
and the system approaches to a quasi-1D magnet with frustrated inter-chain interactions $J_{1y}$ and $J_2$. 
This system has been theoretically studied and it is known that a dimer order or ($Z_2$) quantum spin liquid state appears 
when the frustration between two inter-chain couplings $J_{1y}$ and $J_2$ is quite strong~\cite{Starykh2004, Jiang2012, Metavitsiadis2014}. 
Therefore, we can draw the ground-state phase diagram under an electric field as in Fig.~\ref{frust} (e). 
Namely, a sufficiently strong electric field is expected to create a spin liquid state similarly to the case of the triangular lattice. 
In the case of $\theta_E = \pi/4$, 
the system approaches to a triangular AFM Heisenberg model with an additional interaction $J_{2 \perp}$. 
N\'{e}el ordered layered vanadium oxides such as PbVO$_2$\cite{Tsirlin2008, Oka2008} and 
VOMoO$_4$ \cite{Bombardi2005} are good candidate materials for the $J_1$-$J_2$ magnet. 
They have a relatively large value of $J_2/J_1$ and 
thus a small electric field can make the N\'{e}el state change into dimer or spin liquid states. 

\textit{Quasi-One-Dimensional Magnets. ---}
Purely 1D magnets do not show any magnetic orders even at low temperature due to strong fluctuation 
effects~\cite{Gogolin_book, Giamarchi_book, Tsvelik_book, Mermin1966}. 
However, in a broad class of quasi-1D magnets, a magnetically ordered phase generally appears 
due to a weak but finite three dimensionality if temperature is sufficiently low. 

\begin{figure}[t]
\includegraphics[width=8.7cm]{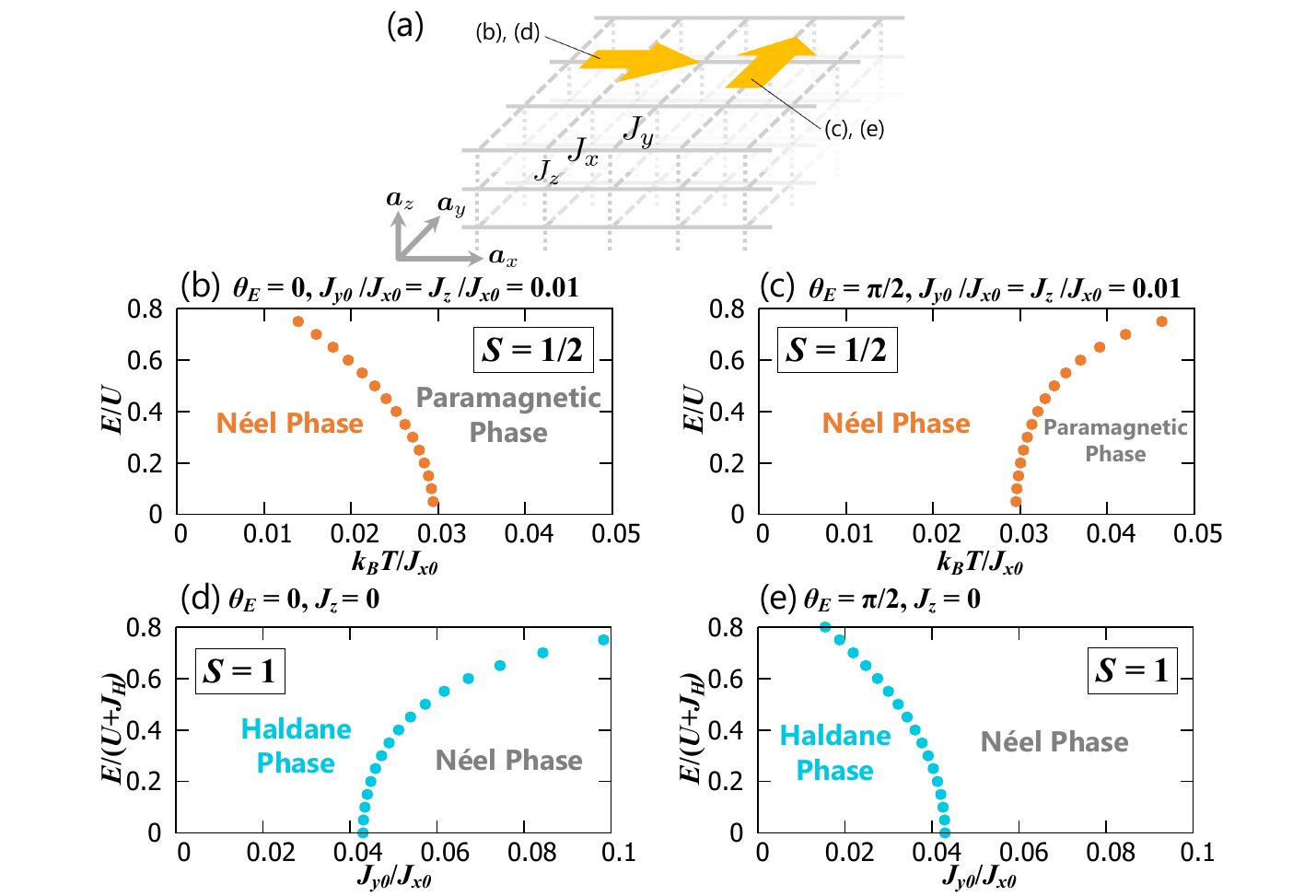}
\caption{(a) Lattice structure of the quasi-one-dimensional AFM Heisenberg magnet consisting of 
weakly coupled 1D Heisenberg chain. Yellow arrows denote the direction of the applied DC electric field. 
(b-e) Phase diagrams of quasi-1D AFM Heisenberg model subject to DC electric fields. 
Panels (b, c) and (d, e) respectively correspond to the spin-$\frac{1}2$ and spin-1 results. 
The electric field $\bm E$ is parallel to the chain ($\bm E \parallel \bm a_x$) in the cases (b,d), while $\bm E$ is 
perpendicular to the chain ($\bm E \parallel \bm a_y$) in the cases (c,e).}
\label{PD}
\end{figure}

When the electric field is parallel to the chain direction, the one dimensionality is further enhanced and 
an exotic quantum phases should appear. On the other hand, an electric field perpendicular to the chain 
makes the inter-chain coupling stronger and the system is expected to show a magnetic long-range order. 
To demonstrate our proposal, we analyze an AFM Heisenberg model on a cubic lattice 
consisting of weakly coupled spin chains, which is depicted in Fig.~\ref{PD} (a). 
In this model, the spin chains are parallel to the $x$ direction, a DC electric field $\bm E$ is in the $x$-$y$ plane, 
and the direction of $\bm E$ is defined by the angle $\theta_E$. 
In this setup, the spin Hamiltonian is written as
\begin{align}
\calH &= \sum_{\bm r} [ J_x(E, \theta_E) \bm S_{\bm r} \cdot \bm S_{\bm r + \bm a_x} 
+ J_y(E, \theta_E) \bm S_{\bm r} \cdot \bm S_{\bm r + \bm a_y}] \nonumber\\
 & \quad+ \sum_{\bm r}  J_z \bm S_{\bm r} \cdot \bm S_{\bm r + \bm a_z}.
\end{align}
For the spin-$\frac{1}{2}$ case, the exchange couplings on the $x$ and $y$ directions are respectively given by 
$J_x(E,\theta_E)=J_{x0}/[1-\{E\cos \theta_E/U\}^2]$, 
$J_y(E,\theta_E)=J_{y0}/[1-\{E\cos (\theta_E-\pi/2)/U\}^2]$, and $J_z$ is that of the $z$ direction. 
For the spin-1 case, the above formulas of the exchange couplings are modified 
by the replacement $U \rightarrow U + J_H$. 

In the spin-$\frac{1}{2}$ system, if temperature becomes low enough (typically, order of inter-chain couplings), 
a N\'{e}el ordered phase emerges. In general, various sorts of finite-temperature phase transition points 
in quasi-1D systems can be determined by applying the chain mean field theory 
(MFT)~\cite{Scalapino1975, Schulz1996, Bocquet2001, Sato2004, Okunishi2007, Klanjsek2008, Sato2013}. 
In fact, the transition points predicted by chain MFT quite agree with 
experimental results of some quasi-1D magnets~\cite{Okunishi2007, Klanjsek2008}. 
We apply the chain MFT to the present spin-$\frac{1}{2}$ system and the resultant phase diagrams on the plane 
$(k_BT,E)$ are summarized in Fig.~\ref{PD} (b) and (c).  The detail of the chain MFT is explained in Supplementary Material. 
The phase diagrams show that when $\bm E$ is parallel to the $x$ ($y$) direction, 
the one dimensionality is enhanced (inter-chain interaction becomes stronger) and the transition temperature decreases 
(grows) with increasing $E$. 

For the spin-$1$ case, the so-called Haldane phase, a typical symmetry-protected-topological phase, is realized 
in each spin-$1$ AFM chain in a parameter range with small inter-chain couplings, 
while a N\'eel ordered phase takes place when inter-chain couplings are strong enough. 
The low-energy properties of the quasi-1D spin-$1$ system have been accurately 
investigated~\cite{Matsumoto2001, Yasuda2005, Sato2007} 
and a quantum Monte Carlo simulation~\cite{Matsumoto2001} shows the quantum phase transition 
between Haldane and N\'eel phases is located at $J_y(E,\theta_E)\simeq0.043J(E,\theta_E)$
for the 2D limit with $J_z=0$. 
Using this relation, we can generally determine the ground-state phase diagram of 
the spatially anisotropic 2D spin-$1$ magnets under an electirc field $\bm E$. 
Figure~\ref{PD} (d) and (e) are respectively the phase diagrams for $\bm E\parallel \bm a_x$ ($\theta_E=0$) and 
$\bm E \parallel \bm a_y$ ($\theta_E=\pi/2$). 
The results of Fig.~\ref{PD} clearly indicate that we can create/annihilate ordered or topological phases of 
quasi-1D magnets with a sufficiently strong DC electric field. 

Our predictions of Fig.~\ref{PD} are generally relevant to a wide class of quasi-1D magnets. 
For example, Sr$_2$CuO$_3$~\cite{Schlappa2012}, Cs$_2$CuCl$_4$~\cite{Kohno2007, Coldea2002}, 
KCuF$_3$~\cite{Mourigal2013} and NMP-TCNQ~\cite{Epstein1971}
(NENP~\cite{Renard1987, Katsumata1989}
, TMNIN~\cite{Gadet1991}
and Y$_2$BaNiO$_5$~\cite{Darriet1993, Sakaguchi1996}) are well known 
as typical quasi-1D spin-$\frac{1}{2}$ (spin-1) magnets. 
Particularly, the Coulomb interaction of NMP-TCNQ has been estimated as 
a rather small $U \sim 0.17 \mathrm{eV}$~\cite{Epstein1971}. 
For this magnet, Fig.~\ref{PD}(c) predicts that the critical temperature can increase by about 50$\%$ 
if we apply DC fields $E^{\parallel}\sim0.8\mathrm{MV/cm}$ along the interchain direction 
with lattice constant $\sim15\AA$. 

\textit{Summary. ---}
In this Letter, we have shown that DC electric fields can enhance the AFM Heisenberg coupling in general Mott insulators 
[See Eqs.~(\ref{eq:spin-1/2}) and (\ref{eq:spin-1})]. 
Then we have illustrated that this enhancement is very useful for controlling the phases of magnets, 
and given rich phase diagrams (See Figs.~\ref{tri}-\ref{PD}). 
We emphasize that a weaker DC field than that for the Mott breakdown is sufficient to control the magnetism, 
and our method is basically free from heating issues in contrast with the AC-field control.  

\begin{acknowledgments}
We would like to thank Takashi Oka for fruitful discussions at the early stage of this work. 
We also thank Hironori Yamaguchi, Toshiya Ideue, Yoshihiro Iwasa and Norio Kawakami for very helpful comments.
K. T. is supported by JSPS KAKENHI (Grant No. JP16J05078) and a JPSJ Research Fellowship for Young Scientists. 
M. S. is supported by Grant-in-Aid for Scientific Research on Innovative Area, Nano Spin Conversion Science
(Grant No. 17H05174), and JSPS KAKENHI (Grant Nos. JP17K05513 and JP15H02117). 
\end{acknowledgments}

\clearpage
\appendix
\renewcommand{\theequation}{S\arabic{equation}}
\setcounter{equation}{0}
\renewcommand{\thefigure}{S\arabic{figure}}
\setcounter{figure}{0}
\renewcommand{\theenumiv}{S\arabic{enumiv}}

\begin{widetext}
\begin{center}
 \textbf{\Large Supplemental Material: Control of Magnetic and Topological Orders with a DC Electric Field}
\end{center}

\subsection*{S1. Derivation of the spin-$\frac{1}{2}$ effective model from a single-band Hubbard model under a DC field}

This section is devoted to the derivation of the spin-$\frac{1}{2}$ effective model~(2). 
We start from a half-filled, repulsive Hubbard model ($U>0$) with an arbitrary on-site potential term. 
The Hamiltonian reads
\begin{align}
\calH &= \sum_{ij \sigma}  t_{ij} c^\dagger_{i \sigma} c_{j \sigma}  + U \sum_i n_{i \uparrow} n_{i \downarrow} 
+ \sum_{i \sigma} V_i n_{i \sigma} \nonumber\\
&= \calH_t + \calH_U + \calH_V \label{Model_Onsite},
\end{align}
where $\calH_t$, $\calH_U$, and $\calH_t$ denote the electron hopping, the on-site Coulomb interaction, 
and the on-site potential, respectively. We assume that all the on-site potential energies are smaller than 
the Coulomb interaction energy, i.e., $|V_i| < U$. 

In order to perform perturbative calculations for any quantum system, 
it is generally useful to introduce projection operators onto Hilbert subspaces.
Let us divide the full Hilbert space into a low- and high-energy states, 
$\{\ket{\Psi_g}\}$ and $\{\ket{\Psi_e}\}$, 
and define the projection operator onto the low-energy (high-energy) state $P_e$ ($P_g$). 
Using these instruments, we can arrive at the effective Hamiltonian for the low-energy subspace 
in the second-order perturbation theory: 
\begin{align}
\calH_\eff = \calH_{gg} + \calH_{ge} \frac{1}{E_g - \calH_{ee}} \calH_{eg}, 
\label{Formula_Heff}
\end{align}
where $\calH_{\alpha \beta} = P_\alpha \calH P_\beta$ $(\alpha, \beta = g, e)$ and 
$E_g$ is defined by $\calH_{gg} \ket{\Psi_g} = E_g \ket{\Psi_g}$. 

We apply the above formula (\ref{Formula_Heff}) to the Mott insulating state of the Hubbard model (\ref{Model_Onsite}) in the strong-coupling limit, i.e. $U \to \infty$. 
In this limit, the ground states of the unperturbed Hamiltonian $\calH_U + \calH_V$ are states where all sites are singly occupied.  
We treat the hopping term $\calH_t$ as the perturbation, and define the low-energy (high-energy) subspace as the ground states (states with doubly occupied sites). 

First we consider $\calH_{eg}$ in the Mott insulating state of the Hubbard model~(\ref{Model_Onsite}). 
In the three terms $\calH_t$, $\calH_U$ and $\calH_V$ of the Hamiltonian $\calH$, only the hopping $\calH_t$ has 
a matrix element between high and low-energy states, $\{\ket{\Psi_g}\}$ and $\{\ket{\Psi_e}\}$. 
Therefore $\calH_{eg}$ is written as 
\begin{align}
\calH_{eg}&=P_e \left( \sum_{ij \sigma}  t_{ij} c^\dagger_{i \sigma} c_{j \sigma} \right) P_g.
\end{align}
From the Pauli's exclusion principle and the half-filled condition, we see that $\calH_{eg}$ survives 
only when the spin indices $\sigma$ on the $i$-th and $j$-th sites are different, i.e., 
$(S^z_i, S^z_j) = (\uparrow, \downarrow), (\downarrow,\uparrow)$, as shown in Fig.~\ref{process}~(a). 
We may thus rewrite $\calH_{eg}$ as  
\begin{align}
\calH_{eg}&= \sum_{ij \sigma}  t_{ij} c^\dagger_{i \sigma} c_{j \sigma} (S^z_i - S^z_j)^2.
\end{align}
Next we compute the energy difference between the ground and the intermediate high-energy states 
depicted in Fig.~\ref{process}. 
To this end, we may focus on two neighboring $i$-th and $j$-th sites. 
In the ground states, both the sites are singly occupied and thus their energy is given by $V_i + V_j$. 
In contrast, the $i$-th site is doubly occupied and the $j$-th site is vacant in the intermediate states. 
Thus the energy is $U + 2V_i$. These results lead to 
\begin{align}
\frac{1}{E_g - \calH_{ee}} \calH_{eg} &=  
\sum_{ij \sigma} \frac{1}{(V_i + V_j) - (U + 2V_i)}  t_{ij} c^\dagger_{i \sigma} c_{j \sigma} (S^z_i - S^z_j)^2 \nonumber \\
&= -\sum_{ij \sigma}\frac{1}{U - \Delta V_{ij}}   t_{ij} c^\dagger_{i \sigma} c_{j \sigma} (S^z_i - S^z_j)^2. \label{eq:inter}
\end{align}
Here we define $\Delta V_{ij} = V_i  -V_j$.

\begin{figure}[thbp]
\includegraphics[width=10cm]{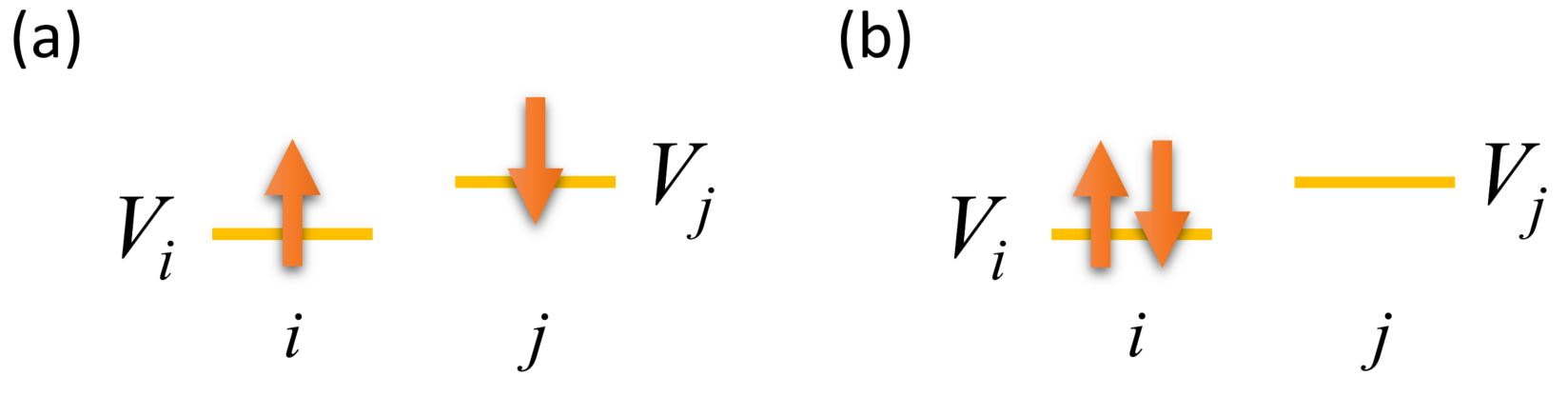}
\caption{Spin configuration of states relevant to the second-order perturbative calculation: 
(a) A ground state and (b) an intermediate state.} 
\label{process}
\end{figure}

Finally, we operate the $\calH_{ge}$ to Eq.~(\ref{eq:inter}) and 
then the second-order perturbation term is calculated as follows: 
\begin{align}
\calH_{ge} \frac{1}{E_g - \calH_{ee}} \calH_{eg} &= - P_g \sum_{i^\prime j^\prime \sigma^\prime}  t_{j^\prime i^\prime} c^\dagger_{j^\prime \sigma^\prime} c_{i^\prime \sigma^\prime}\sum_{ij \sigma}\frac{1}{U - \Delta V_{ij}}   t_{ij} c^\dagger_{i \sigma} c_{j \sigma} (S^z_i - S^z_j)^2 \nonumber\\
&= - \sum_{ij \sigma} \frac{|t_{ij}|^2}{U - \Delta V_{ij}} c^\dagger_{j \sigma} c_{i \sigma}  c^\dagger_{i \sigma} c_{j \sigma} (S^z_i - S^z_j)^2  - \sum_{ij \sigma} \frac{|t_{ij}|^2}{U - \Delta V_{ij}} c^\dagger_{j \bar{\sigma}} c_{i \bar{\sigma}}  c^\dagger_{i \sigma} c_{j \sigma} (S^z_i - S^z_j)^2  \nonumber\\
&= - \sum_{ij \sigma} \frac{|t_{ij}|^2}{U - \Delta V_{ij}} n_{j \sigma} (1-n_{i\sigma}) (S^z_i - S^z_j)^2  + \sum_{ij \sigma} \frac{|t_{ij}|^2}{U - \Delta V_{ij}} c^\dagger_{j \bar{\sigma}} c_{j \sigma}  c^\dagger_{i \sigma}c_{i \bar{\sigma}}  (S^z_i - S^z_j)^2 \nonumber \\
&= - \sum_{ij} \frac{|t_{ij}|^2}{U - \Delta V_{ij}} (S^z_i - S^z_j)^2 + \sum_{ij} \frac{|t_{ij}|^2}{U - \Delta V_{ij}} (S^-_j S^+_i + S^+_j S^-_i) (S^z_i - S^z_j)^2  \nonumber\\
&= - \sum_{ij} \frac{|t_{ij}|^2}{U - \Delta V_{ij}} (S^z_i - S^z_j)^2 + \sum_{ij} \frac{|t_{ij}|^2}{U - \Delta V_{ij}} (S^-_j S^+_i + S^+_j S^-_i) \nonumber\\
&=  \sum_{ij} \frac{|t_{ij}|^2}{U - \Delta V_{ij}} \left( -\frac{1}{2} + 2 S^z_i S^z_j + S^+_i S^-_j + S^-_i S^+_j \right)\nonumber\\
&= \sum_{ij} \frac{2 |t_{ij}|^2}{U - \Delta V_{ij}} \bm S_i \cdot \bm S_j + \mathrm{const.}, \label{eq:final}
\end{align}
where we have defined $\bar{\sigma} = - \sigma$. The first-order term $\calH_{gg}$ gives only a constant term, and 
therefore the effective Hamiltonian up to the second-order perturbation theory is given by  
\begin{align}
\calH_\eff &= \sum_{ij} \frac{2 |t_{ij}|^2}{U - \Delta V_{ij}} \bm S_i \cdot \bm S_j+ \mathrm{const.}\nonumber\\ 
&= \sum_{\langle ij \rangle} \frac{4 |t_{ij}|^2}{U} \frac{1}{1- \left( \frac{\Delta V_{ij}}{U} \right)^2} \bm S_i \cdot \bm S_j+ \mathrm{const.},
\end{align}
where the summation is taken over the every bond $\langle i, j \rangle$ in the last line. 
This is the effective model (2) in the main text.


\subsection*{S2. Derivation of the spin-1 effective model from a two-band Hubbard model under a DC field}

In this section, we show the derivation of the effective spin-1 model (3). 
We start from a half-filled, two-orbital Hubbard model with an additional on-site potential. 
The Hamiltonian consists of three parts of hopping, interaction, and potential terms:
\begin{eqnarray}
\calH&=& \calH_t + \calH_V + \calH_\mathrm{int}. 
\label{eq:spin1Hubbard}
\end{eqnarray}
These terms are given by 
\begin{align}
\calH_t &=\sum_{ij}\sum_{\alpha}\sum_{\sigma} t_{ij} c^\dagger_{i \alpha \sigma} c_{j \alpha \sigma},\\
\calH_V &= \sum_{i \sigma} V_i n_{i \alpha \sigma},\\
\calH_\mathrm{int} 
&=
U \sum_{i}\sum_{\alpha} n_{i \alpha  \uparrow} n_{i \alpha \downarrow} 
+  U^\prime \sum_{i}\sum_{\sigma \sigma^\prime} n_{i 1 \sigma} n_{i 2 \sigma^\prime} \nonumber \\
&\qquad 
- J  \sum_{i} \sum_{\sigma \sigma^\prime} c^\dagger_{i 1 \sigma} c_{i 1 \sigma^\prime } c^\dagger_{i 2 \sigma^\prime}c_{i 2\sigma}
- J_P  \sum_{i} \left( c^\dagger_{i 1 \uparrow} c^\dagger_{i 1 \downarrow } c_{i 2 \downarrow } c_{i 2 \uparrow} + \mathrm{h.c.} \right). \label{eq:twoorbHubint}
\end{align} 
Here $\alpha(=1,2)$ is orbital index and $\bar{\sigma}$ denotes the opposite spin $-\sigma$. In the interaction $\calH_\mathrm{int}$, 
$U$, $U'$, $J$, and $J_P$ terms denote an intra-orbital interaction, an inter-orbital interaction, a Hund's coupling and a pair hopping respectively. Due to the rotational symmetry of Coulomb interaction, $J_P = J$ is required. For convenience, we transform the interaction $\calH_\mathrm{int}$ (\ref{eq:twoorbHubint}) as follows: 
\begin{align}
\calH_\mathrm{int} &=  U \sum_{i} \sum_\alpha n_{i \alpha  \uparrow} n_{i \alpha \downarrow} 
+  \left( U^\prime - \frac{J}{2}\right) \sum_{i}\sum_{\sigma \sigma^\prime} n_{i 1\sigma} n_{i 2\sigma^\prime} 
- 2 J  \sum_{i} \bm S_{i1} \cdot \bm S_{i2} 
- J  \sum_{i} \left( c^\dagger_{i 1 \uparrow} c^\dagger_{i 1 \downarrow } c_{i 2 \downarrow } c_{i 2 \uparrow} + \mathrm{h.c.} \right),
\label{eq:Hund}
\end{align}
where we have used the identity 
\begin{align}
\sum_\sigma c^\dagger_{i 1 \sigma}c_{i 1 \bar{\sigma}} c^\dagger_{i 2 \bar{\sigma}} c_{i 2 \sigma} 
=2 \bm S_{i1} \cdot \bm S_{i2} - \frac{1}{2} \sum_\sigma n_{i1 \sigma} n_{i2 \sigma} 
+ \frac{1}{2} \sum_\sigma n_{i1 \sigma} n_{i2 \bar{\sigma}},
\end{align}
and $\bm S_{i\alpha}$ is the spin operator for an $\alpha$-orbital electron on $i$-th site. 

First we discuss the ground state under the condition of both the half-filling and the strong-coupling 
limit $U > U^\prime > J \gg t$. 
In this condition, all the orbits are singly occupied and there are two electrons per one site in the ground states. 
We here introduce local bases $\ket{\psi}_i$ to represent the spin state on each site $i$. 
They are classified into the spin-triplet sector $\mathcal{T}_i$ and the spin-singlet sector $\mathcal{S}_i$ : 
\begin{align}
\mathcal{T}_i &= \{  \ket{+}_i, \ket{\circ}_i, \ket{-}_i  \}, \\\mathcal{S}_i &= \{ \ket{s}_i \}. 
\end{align}
Four kinds of $\ket{\psi}_i$ are defined as 
\begin{align}
\ket{+}_i &= c^\dagger_{i 1 \uparrow} c^\dagger_{i 2 \uparrow}  \ket{0}, \\ \ket{-}_i &= c^\dagger_{i 1 \downarrow} c^\dagger_{i 2 \downarrow}  \ket{0},\\
\ket{\circ}_i & = \frac{1}{\sqrt{2}} \left( c^\dagger_{i 1 \uparrow} c^\dagger_{i 2 \downarrow} \ket{0} + c^\dagger_{i 1 \downarrow} c^\dagger_{i 2 \uparrow} \ket{0}  \right),\\
\ket{s}_i & = \frac{1}{\sqrt{2}} \left( c^\dagger_{i 1 \uparrow} c^\dagger_{i 2 \downarrow} \ket{0} - c^\dagger_{i 1 \downarrow} c^\dagger_{i 2 \uparrow} \ket{0}  \right),
\end{align}
where $\ket{+}_i$, $\ket{-}_i$, and $\ket{\circ}_i$ are respectively the $S^z=+1$, $-1$, and $0$ state on $i$-th site.  
Within this localized spin subspace, the correlation function of two-orbital spins on single site is computed as 
\begin{align}
\bra{\psi}_i \bm S_{i1} \cdot \bm S_{i2} \ket{\psi}_i = 
\begin{cases}
\frac{1}{4} & (\ket{\psi}_i \in \mathcal{T}_i) \\
-\frac{3}{4} & (\ket{\psi}_i \in \mathcal{S}_i). 
\end{cases}
\end{align}
This result and Hund's coupling in Eq.~(\ref{eq:Hund}) clearly show that the ground state on each site is in the spin-triplet sector, 
namely, localized spin-1 system is realized in Eq.~(\ref{eq:spin1Hubbard}). 

Next, 
we focus on the zero-potential case of $V_i= 0$. As one will see soon later, the effective model for $V_i \neq 0$ 
can be easily derived by simply extending the result of the $V_i= 0$ case. 
Using the formula (\ref{Formula_Heff}), let us derive the effective spin model for the $V_i= 0$ case 
with the hopping $\calH_t$ being the perturbation. 
To this end, we introduce the nine local bases $\ket{\psi_i}_i\ket{\psi_j}_j$
which represent neighboring $i$-th and $j$-th spin states ($\psi_{i,j}\in \{+,\circ,-\}$). 
In the matrix form, the bases are expressed as
\begin{align}
\ket{\Psi_{ij}} =
\begin{pmatrix}
\begin{array}{c}
\ket{+}_i \ket{+}_j \\
\ket{+}_i \ket{\circ}_j \\
\ket{+}_i \ket{-}_j \\
\ket{\circ}_i \ket{+}_j \\
\ket{\circ}_i \ket{\circ}_j \\
\ket{\circ}_i \ket{-}_j \\
\ket{-}_i \ket{+}_j \\
\ket{-}_i \ket{\circ}_j \\
\ket{-}_i \ket{-}_j 
\end{array}
\end{pmatrix}.
\end{align}
Through straightforward calculation, we obtain
\begin{align}
\calH_{ge}\calH_{eg}\ket{+}_i \ket{-}_j&=- \ket{\circ}_i \ket{\circ}_j  + 2 \ket{+}_i \ket{-}_j, \\
\calH_{ge}\calH_{eg} \ket{+}_i \ket{0}_j &= -\ket{0}_i \ket{+}_j + \ket{+}_i \ket{0}_j,\\
\calH_{ge}\calH_{eg} \ket{0}_i \ket{0}_j &= -\ket{+}_i \ket{-}_j - \ket{-}_i \ket{+}_j + \ket{\circ}_i \ket{\circ}_j ,
\end{align}
and
\begin{align}
\calH_{ge}\frac{1}{E_g-\calH_{ee}}\calH_{eg}\ket{+}_i \ket{-}_j&=\frac{|t_{ij}|^2}{\Delta E_{ij}} \left(- \ket{\circ}_i \ket{\circ}_j  + 2 \ket{+}_i \ket{-}_j \right), \\
\calH_{ge}\frac{1}{E_g-\calH_{ee}}\calH_{eg} \ket{+}_i \ket{0}_j &=\frac{|t_{ij}|^2}{\Delta E_{ij}}\left( -\ket{0}_i \ket{+}_j + \ket{+}_i \ket{0}_j \right),\\
\calH_{ge}\frac{1}{E_g-\calH_{ee}}\calH_{eg} \ket{0}_i \ket{0}_j &=\frac{|t_{ij}|^2}{\Delta E_{ij}}\left( -\ket{+}_i \ket{-}_j - \ket{-}_i \ket{+}_j + \ket{\circ}_i \ket{\circ}_j \right),
\end{align}
where
\begin{align}
\Delta E_{ij} &= \left\{ 2 \times \left( U^\prime- \frac{J}{2}\right) - 2 \times 2 J \cdot \frac{1}{4}\right\} - \left\{ U + 2 \times \left( U^\prime- \frac{J}{2}\right) \right\} \nonumber\\
&=-(U+J).
\label{eq:deltaE}
\end{align}
From these results, the effective Hamiltonian in the $i$-th and $j$-th sites is given by
\begin{align}
\bra{\Psi_{ij}} \calH_\eff \ket{\Psi_{ij}}&=
 E_g I + 
\frac{|t_{ij}|^2}{U+J}
\begin{pmatrix}
 0 & 0 & 0 & 0 & 0 & 0 & 0 & 0 & 0 \\
 0 & -1 & 0 & 1 & 0 & 0 & 0 & 0 & 0 \\
 0 & 0 & -2 & 0 & 1 & 0 & 0 & 0 & 0 \\
 0 & 1 & 0 & -1 & 0 & 0 & 0 & 0 & 0 \\
 0 & 0 & 1 & 0 & -1 & 0 & 1 & 0 & 0 \\
 0 & 0 & 0 & 0 & 0 & -1 & 0 & 1 & 0 \\
 0 & 0 & 0 & 0 & 1 & 0 & -2 & 0 & 0 \\
 0 & 0 & 0 & 0 & 0 & 1 & 0 & -1 & 0 \\
 0 & 0 & 0 & 0 & 0 & 0 & 0 & 0 & 0
\end{pmatrix}. \label{eq:matele}
\end{align}
On the other hand, the matrix elements of Heisenberg interaction between two spin-1 operators are computed as 
\begin{align}
\bra{\Psi_{ij}} \bm S_{i} \cdot \bm S_{j} \ket{\Psi_{ij}}&=
\begin{pmatrix}
 1 & 0 & 0 & 0 & 0 & 0 & 0 & 0 & 0 \\
 0 & 0 & 0 & 1 & 0 & 0 & 0 & 0 & 0 \\
 0 & 0 & -1 & 0 & 1 & 0 & 0 & 0 & 0 \\
 0 & 1 & 0 & 0 & 0 & 0 & 0 & 0 & 0 \\
 0 & 0 & 1 & 0 & 0 & 0 & 1 & 0 & 0 \\
 0 & 0 & 0 & 0 & 0 & 0 & 0 & 1 & 0 \\
 0 & 0 & 0 & 0 & 1 & 0 & -1 & 0 & 0 \\
 0 & 0 & 0 & 0 & 0 & 1 & 0 & 0 & 0 \\
 0 & 0 & 0 & 0 & 0 & 0 & 0 & 0 & 1 
\end{pmatrix},
\label{eq:spin-1_mat}
\end{align}
Comparing Eqs.~(\ref{eq:matele}) and (\ref{eq:spin-1_mat}), we see the identity
\begin{align}
\bra{\Psi_{ij}} \calH_\eff \ket{\Psi_{ij}}&= E_g I + 
\frac{|t_{ij}|^2}{U-J}
\left( \bra{\Psi_{ij}} \bm S_{i} \cdot \bm S_{j} \ket{\Psi_{ij}} - I \right).
\label{Heff_matrixele}
\end{align}
Therefore, without the constant in Eq.~(\ref{Heff_matrixele}), the effective spin model for the zero-potential system is written as
\begin{align}
\calH_\eff &= \sum_{ij} \frac{|t_{ij}|^2}{U + J} \bm S_i \cdot \bm S_j \nonumber\\
&=\sum_{\langle ij \rangle} \frac{ 2|t_{ij}|^2}{U + J} \bm S_i \cdot \bm S_j \nonumber\\
&=J^{S=1}_{ij} \sum_{\langle ij \rangle} \bm S_i \cdot \bm S_j ,
\end{align}
where we have defined $J^{S=1}_{ij} = 2|t_{ij}|^2 / (U + J)$. 

Finally, let us turn to a generic case with $V_i \neq 0$. In this case, most of the perturbative calculations are
the same as those of the $V_i=0$ case. However, $\Delta E_{ij}$ of Eq.~(\ref{eq:deltaE}) should be changed into 
\begin{align}
\Delta E_{ij} &= \left\{ 2 \times \left( U^\prime- \frac{J}{2} \right) - 2 \times 2 J \cdot \frac{1}{4}\right) + 2V_i +2 V_j \} 
- \left\{ U + 2 \times \left( U^\prime- \frac{J}{2}\right)  + 3 V_i  + V_j \right\} \nonumber\\
&=-(U+J + \Delta V_{ij}),
\end{align}
where $\Delta V_{ij} = V_i -V_j$. 
Thus the effective spin model for the two-orbital Hubbard model with an on-site potential is written as
\begin{align}
\calH_\eff &= \sum_{ij} \frac{|t_{ij}|^2}{U + J + \Delta V_{ij}} \bm S_i \cdot \bm S_j \nonumber\\
&=\sum_{\langle ij \rangle} \frac{ J^{S=1}_{ij}}{1 - \left(\frac{\Delta V_{ij}}{U+J}\right)^2}  \bm S_i \cdot \bm S_j.
\end{align}
This is the effective model (3) in the main text.

\subsection*{S3. Bosonization and Chain Mean-field Approach} 
In this section, we shortly explain the computation process of the critical temperature 
between N\'eel ordered and paramagnetic phases in our quasi-1D spin-$\frac{1}{2}$ model (5). 
First we summarize some results of the bosonization for spin-$\frac{1}{2}$ chains~[44, 47-49].
Then, by combining the chain mean-field theory (MFT) with the bosonization results~[83-89].
we determine the critical temperature of the quasi-1D model (5). 

We start from the definition of the 1D spin-$\frac{1}{2}$ XXZ chain model. 
The Hamiltonian is given by  
\begin{eqnarray}
{\cal H}_{\rm xxz}&=& J\sum_j \left[S^x_{j}S^x_{j+1}+S^y_{j}S^y_{j+1}+\Delta_zS^z_{j}S^z_{j+1}\right]
-H\sum_jS^z_j,
\end{eqnarray}
where $\bm S_j$ is the spin-$\frac{1}{2}$ operator in $j$-th site, $J>0$ is the strength of the exchange interaction, 
$\Delta_z$ is the XXZ anisotropy parameter, and $H$ is the external magnetic field. 
The point of $\Delta_z=1$ and $H=0$ corresponds to the SU(2)-symmetric antiferromagnetic Heisenberg model. 
The XXZ model is a typical integrable system and the TL-liquid phase with gapless spinon excitations widely exists 
in the range $-1 < \Delta_z \leq 1$ at zero field $H=0$. The TL liquid phase survives from zero field to the saturation field.  
The bosonization can accurately describe the low-energy properties in/around the TL-liquid phase. 
Through the standard bosonization process, the XXZ model in/around the TL-liquid phase 
is mapped to a low-energy gapless scalar-field theory, whose Hamiltonian is 
\begin{eqnarray}
{\cal H}_{\rm eff}=\int dx\,\, \frac{v}{2}\,
\Big[\frac{1}{K}(\partial_x\phi)^2+K(\partial_x\theta)^2\Big],
\label{eq:eff}
\end{eqnarray}
where $x=j a_0$ is the continuous coordinate ($a_0$ : lattice constant), and 
$\phi(x,t)$ and $\theta(x,t)$ are the canonical pair of real scalar fields satisfying the commutation relation 
$[\phi(x,t),\partial_y\theta(y,t)]=i\delta(x-y)$. 
Two symbols $v$ and $K$ respectively denote the spinon group velocity and the TL-liquid parameter. 
For instance, $K=1$ and $v=\pi J a_0/2$ at the SU(2) point. 
Spin operators are also bosonized as 
\begin{eqnarray}
S_{j}^z &\approx& M\,+\,\frac{a_0}{\sqrt{2\pi}}\partial_x{\phi}
+(-1)^j A_1\cos\left(\sqrt{2\pi}{\phi}
 + 2\pi Mj\right)+\cdots,\nonumber\\
S_{j}^+  &\approx& e^{i\sqrt{2\pi}\theta}\left[(-1)^j B_0
+B_1\cos\left(\sqrt{2\pi}{\phi}
 + 2\pi Mj\right)+\cdots\right]. 
\label{eq:spin_boson}
\end{eqnarray}
where 
$M=\langle S_j^z\rangle$ is the $H$-induced uniform magnetization per one site, 
and $A_n$ and $B_n$ are non-universal constants depending on the model parameters $J$, $\Delta_z$ and $H$. 
The accurate values of $v$, $K$, $A_n$ and $B_n$ have been computed by using Bethe ansatz and numerical methods~[103-108].
On the basis of the formulas (\ref{eq:eff}) and (\ref{eq:spin_boson}), 
one can correctly calculate the long-distance or long-time behavior of correlation functions in the TL-liquid phase.
Let us define the dynamical spin susceptibility with the wave number $k$ and frequency $\omega$ as  
$\chi_R^{ab}(k,\omega)=-\int_{0}^\beta d\tau \sum_k e^{-ik j a_0 +i\omega_n \tau}
\langle T_\tau S^a_j (\tau)S^b_0(0)\rangle|_{i\omega_n=\omega+i\eta}$, where $\tau$ is imaginary time, 
$\beta=1/(k_BT)$ is inverse temperature, $\omega_n=2\pi n/\beta$ ($n$: integer), and $\eta$ is an infinitesimal positive constant. 
Through the bosonization technique with Eqs.~(\ref{eq:eff}) and (\ref{eq:spin_boson}), 
one can calculate the transverse dynamical susceptibility around $k=\pi+\delta k$ 
in the TL-liquid phase: 
\begin{eqnarray}
\chi_R^{-+}(\pi+\delta k,\omega) &\approx&
 -B_0^2 \,\,\frac{a_0}{v}\,\,\sin\Big(\frac{\pi}{2K}\Big)\,\,
\Big(\frac{2\pi a_0}{\beta v}\Big)^{1/K-2}\nonumber\\
&&\times
B\Big(-i\frac{\beta(\omega-v \delta k)}{4\pi}+\frac{1}{4K},1-\frac{1}{2K}\Big)
B\Big(-i\frac{\beta(\omega+v \delta k)}{4\pi}+\frac{1}{4K},1-\frac{1}{2K}\Big),
\label{eq:transverse}
\end{eqnarray}
where $S^\pm_j=S^x_j\pm i S^y_j$ and $B(x,y)$ is Beta function. 
This formula is quite reliable in the range of $|\delta k|\ll a_0^{-1}$ and $|\omega|\ll J, k_BT$. 
In the TL-liquid phase of the XXZ chain, the relation 
$\chi_R^{xx}(k,\omega)=\chi_R^{yy}(k,\omega)=\frac{1}{2}\chi_R^{-+}(k,\omega)$ holds. 

Next, we apply the chain MFT to our quasi-1D spin-$\frac{1}{2}$ magnet (5) with the above bosonization results.  
In the chain MFT, we accurately take into account quantum and thermal fluctuation effects in the strong coupled 1D direction, 
while an inter-chain interaction is treated within the standard MFT. On the basis of this approach, 
the $S^x$ component of the dynamical spin susceptibility in the quasi-1D system (5) is calculated 
as the following RPA-like form: 
\begin{eqnarray}
\chi_{\rm 3D}^{xx}(k_x,k_y,k_z,\omega) &=& \frac{\chi_{R}^{xx}(k_x,\omega)}
{1-2(J_y\cos k_y+J_z\cos k_z)\chi_{R}^{xx}(k_x,\omega)},
\label{eq:XXZchain-MFT}
\end{eqnarray}
where the wave number $k_x$ corresponds to the 1D-chain direction, and $k_{y,z}$ are the wave numbers 
along the inter-chain direction. This result is quantitatively valid in the sufficiently weak inter-chain regime $|J_{y,z}|\ll J$.  
The phase transition between the N\'eel and paramagnetic phases is determined as the point where 
$\chi_{\rm 3D}^{xx}(\pi,\pi,\pi,\omega\to0)$ diverges. This point is equivalent to the condition that the denominator of 
Eq.~(\ref{eq:XXZchain-MFT}) becomes zero at $\bm k=(\pi,\pi,\pi)$ and $\omega\to 0$: 
\begin{eqnarray}
-2(J_y+J_z)\chi_{R}^{xx}(\pi,\omega\to0)=1
\label{eq:chain-MFT_3}
\end{eqnarray}
Substituting the bosonization result (\ref{eq:transverse}) into this condition, we arrive at the formula of 
determining the phase transition temperature: 
\begin{eqnarray}
B_0^2 \,\,\frac{(J_y+J_z) a_0}{v}\,\,\sin\Big(\frac{\pi}{2K}\Big)\,\,
\Big(\frac{2\pi a_0}{\beta v}\Big)^{1/K-2}\,\,B\Big(\frac{1}{4K},1-\frac{1}{2K}\Big)^2=1. 
\label{eq:chain-MFT_4}
\end{eqnarray}
Using this result, we have drawn the phase boundary of Fig.4 (b) and (c). We finally note a technical issue 
that since the parameter $B_0$ is ill-defined just on the SU(2) point of $\Delta_z=1$ and $H=M=0$, 
we have used its value for a nearly SU(2)-symmetric model with an infinitesimal small magnetization $M=0.01$ in Fig. 4. 
\end{widetext}


\begin{thebibliography}{103}%
\makeatletter
\providecommand \@ifxundefined [1]{%
 \@ifx{#1\undefined}
}%
\providecommand \@ifnum [1]{%
 \ifnum #1\expandafter \@firstoftwo
 \else \expandafter \@secondoftwo
 \fi
}%
\providecommand \@ifx [1]{%
 \ifx #1\expandafter \@firstoftwo
 \else \expandafter \@secondoftwo
 \fi
}%
\providecommand \natexlab [1]{#1}%
\providecommand \enquote  [1]{``#1''}%
\providecommand \bibnamefont  [1]{#1}%
\providecommand \bibfnamefont [1]{#1}%
\providecommand \citenamefont [1]{#1}%
\providecommand \href@noop [0]{\@secondoftwo}%
\providecommand \href [0]{\begingroup \@sanitize@url \@href}%
\providecommand \@href[1]{\@@startlink{#1}\@@href}%
\providecommand \@@href[1]{\endgroup#1\@@endlink}%
\providecommand \@sanitize@url [0]{\catcode `\\12\catcode `\$12\catcode
  `\&12\catcode `\#12\catcode `\^12\catcode `\_12\catcode `\%12\relax}%
\providecommand \@@startlink[1]{}%
\providecommand \@@endlink[0]{}%
\providecommand \url  [0]{\begingroup\@sanitize@url \@url }%
\providecommand \@url [1]{\endgroup\@href {#1}{\urlprefix }}%
\providecommand \urlprefix  [0]{URL }%
\providecommand \Eprint [0]{\href }%
\providecommand \doibase [0]{http://dx.doi.org/}%
\providecommand \selectlanguage [0]{\@gobble}%
\providecommand \bibinfo  [0]{\@secondoftwo}%
\providecommand \bibfield  [0]{\@secondoftwo}%
\providecommand \translation [1]{[#1]}%
\providecommand \BibitemOpen [0]{}%
\providecommand \bibitemStop [0]{}%
\providecommand \bibitemNoStop [0]{.\EOS\space}%
\providecommand \EOS [0]{\spacefactor3000\relax}%
\providecommand \BibitemShut  [1]{\csname bibitem#1\endcsname}%
\let\auto@bib@innerbib\@empty
\bibitem [{\citenamefont {Oka}\ and\ \citenamefont {Aoki}(2009)}]{Oka2009}%
  \BibitemOpen
  \bibfield  {author} {\bibinfo {author} {\bibfnamefont {T.}~\bibnamefont
  {Oka}}\ and\ \bibinfo {author} {\bibfnamefont {H.}~\bibnamefont {Aoki}},\
  }\href {\doibase 10.1103/PhysRevB.79.081406} {\bibfield  {journal} {\bibinfo
  {journal} {Phys. Rev. B}\ }\textbf {\bibinfo {volume} {79}},\ \bibinfo
  {pages} {081406} (\bibinfo {year} {2009})}\BibitemShut {NoStop}%
\bibitem [{\citenamefont {Lindner}\ \emph {et~al.}(2011)\citenamefont
  {Lindner}, \citenamefont {Refael},\ and\ \citenamefont
  {Galitski}}]{Lindner2011}%
  \BibitemOpen
  \bibfield  {author} {\bibinfo {author} {\bibfnamefont {N.~H.}\ \bibnamefont
  {Lindner}}, \bibinfo {author} {\bibfnamefont {G.}~\bibnamefont {Refael}}, \
  and\ \bibinfo {author} {\bibfnamefont {V.}~\bibnamefont {Galitski}},\ }\href
  {\doibase 10.1038/nphys1926} {\bibfield  {journal} {\bibinfo  {journal} {Nat.
  Phys.}\ }\textbf {\bibinfo {volume} {7}},\ \bibinfo {pages} {490} (\bibinfo
  {year} {2011})}\BibitemShut {NoStop}%
\bibitem [{\citenamefont {Grushin}\ \emph {et~al.}(2014)\citenamefont
  {Grushin}, \citenamefont {G\'omez-Le\'on},\ and\ \citenamefont
  {Neupert}}]{Grushin2014}%
  \BibitemOpen
  \bibfield  {author} {\bibinfo {author} {\bibfnamefont {A.~G.}\ \bibnamefont
  {Grushin}}, \bibinfo {author} {\bibfnamefont {A.}~\bibnamefont
  {G\'omez-Le\'on}}, \ and\ \bibinfo {author} {\bibfnamefont {T.}~\bibnamefont
  {Neupert}},\ }\href {\doibase 10.1103/PhysRevLett.112.156801} {\bibfield
  {journal} {\bibinfo  {journal} {Phys. Rev. Lett.}\ }\textbf {\bibinfo
  {volume} {112}},\ \bibinfo {pages} {156801} (\bibinfo {year}
  {2014})}\BibitemShut {NoStop}%
\bibitem [{\citenamefont {{Sato}}\ \emph {et~al.}(2014)\citenamefont {{Sato}},
  \citenamefont {{Sasaki}},\ and\ \citenamefont {{Oka}}}]{Sato2014}%
  \BibitemOpen
  \bibfield  {author} {\bibinfo {author} {\bibfnamefont {M.}~\bibnamefont
  {{Sato}}}, \bibinfo {author} {\bibfnamefont {Y.}~\bibnamefont {{Sasaki}}}, \
  and\ \bibinfo {author} {\bibfnamefont {T.}~\bibnamefont {{Oka}}},\ }\href
  {http://arxiv.org/abs/1404.2010} {\bibfield  {journal} {\bibinfo  {journal}
  {arXiv : 1404.2010}\ } (\bibinfo {year} {2014})}\BibitemShut {NoStop}%
\bibitem [{\citenamefont {Wang}\ \emph {et~al.}(2013)\citenamefont {Wang},
  \citenamefont {Steinberg}, \citenamefont {Jarillo-Herrero},\ and\
  \citenamefont {Gedik}}]{Wang2013}%
  \BibitemOpen
  \bibfield  {author} {\bibinfo {author} {\bibfnamefont {Y.~H.}\ \bibnamefont
  {Wang}}, \bibinfo {author} {\bibfnamefont {H.}~\bibnamefont {Steinberg}},
  \bibinfo {author} {\bibfnamefont {P.}~\bibnamefont {Jarillo-Herrero}}, \ and\
  \bibinfo {author} {\bibfnamefont {N.}~\bibnamefont {Gedik}},\ }\href
  {\doibase 10.1126/science.1239834} {\bibfield  {journal} {\bibinfo  {journal}
  {Science}\ }\textbf {\bibinfo {volume} {342}},\ \bibinfo {pages} {453}
  (\bibinfo {year} {2013})}\BibitemShut {NoStop}%
\bibitem [{\citenamefont {Jotzu}\ \emph {et~al.}(2014)\citenamefont {Jotzu},
  \citenamefont {Messer}, \citenamefont {Desbuquois}, \citenamefont {Lebrat},
  \citenamefont {Uehlinger}, \citenamefont {Greif},\ and\ \citenamefont
  {Esslinger}}]{Jotzu2014}%
  \BibitemOpen
  \bibfield  {author} {\bibinfo {author} {\bibfnamefont {G.}~\bibnamefont
  {Jotzu}}, \bibinfo {author} {\bibfnamefont {M.}~\bibnamefont {Messer}},
  \bibinfo {author} {\bibfnamefont {R.}~\bibnamefont {Desbuquois}}, \bibinfo
  {author} {\bibfnamefont {M.}~\bibnamefont {Lebrat}}, \bibinfo {author}
  {\bibfnamefont {T.}~\bibnamefont {Uehlinger}}, \bibinfo {author}
  {\bibfnamefont {D.}~\bibnamefont {Greif}}, \ and\ \bibinfo {author}
  {\bibfnamefont {T.}~\bibnamefont {Esslinger}},\ }\href {\doibase
  10.1038/nature13915} {\bibfield  {journal} {\bibinfo  {journal} {Nature}\
  }\textbf {\bibinfo {volume} {515}},\ \bibinfo {pages} {237} (\bibinfo {year}
  {2014})}\BibitemShut {NoStop}%
\bibitem [{\citenamefont {Takayoshi}\ \emph
  {et~al.}(2014{\natexlab{a}})\citenamefont {Takayoshi}, \citenamefont {Aoki},\
  and\ \citenamefont {Oka}}]{Takayoshi2014a}%
  \BibitemOpen
  \bibfield  {author} {\bibinfo {author} {\bibfnamefont {S.}~\bibnamefont
  {Takayoshi}}, \bibinfo {author} {\bibfnamefont {H.}~\bibnamefont {Aoki}}, \
  and\ \bibinfo {author} {\bibfnamefont {T.}~\bibnamefont {Oka}},\ }\href
  {\doibase 10.1103/PhysRevB.90.085150} {\bibfield  {journal} {\bibinfo
  {journal} {Phys. Rev. B}\ }\textbf {\bibinfo {volume} {90}},\ \bibinfo
  {pages} {085150} (\bibinfo {year} {2014}{\natexlab{a}})}\BibitemShut
  {NoStop}%
\bibitem [{\citenamefont {Takayoshi}\ \emph
  {et~al.}(2014{\natexlab{b}})\citenamefont {Takayoshi}, \citenamefont {Sato},\
  and\ \citenamefont {Oka}}]{Takayoshi2014b}%
  \BibitemOpen
  \bibfield  {author} {\bibinfo {author} {\bibfnamefont {S.}~\bibnamefont
  {Takayoshi}}, \bibinfo {author} {\bibfnamefont {M.}~\bibnamefont {Sato}}, \
  and\ \bibinfo {author} {\bibfnamefont {T.}~\bibnamefont {Oka}},\ }\href
  {\doibase 10.1103/PhysRevB.90.214413} {\bibfield  {journal} {\bibinfo
  {journal} {Phys. Rev. B}\ }\textbf {\bibinfo {volume} {90}},\ \bibinfo
  {pages} {214413} (\bibinfo {year} {2014}{\natexlab{b}})}\BibitemShut
  {NoStop}%
\bibitem [{\citenamefont {Mentink}\ \emph {et~al.}(2015)\citenamefont
  {Mentink}, \citenamefont {Balzer},\ and\ \citenamefont
  {Eckstein}}]{Mentink2015}%
  \BibitemOpen
  \bibfield  {author} {\bibinfo {author} {\bibfnamefont {J.~H.}\ \bibnamefont
  {Mentink}}, \bibinfo {author} {\bibfnamefont {K.}~\bibnamefont {Balzer}}, \
  and\ \bibinfo {author} {\bibfnamefont {M.}~\bibnamefont {Eckstein}},\ }\href
  {http://dx.doi.org/10.1038/ncomms7708} {\bibfield  {journal} {\bibinfo
  {journal} {Nature Communications}\ }\textbf {\bibinfo {volume} {6}},\
  \bibinfo {pages} {6708 EP } (\bibinfo {year} {2015})},\ \bibinfo {note}
  {article}\BibitemShut {NoStop}%
\bibitem [{\citenamefont {Sato}\ \emph {et~al.}(2016)\citenamefont {Sato},
  \citenamefont {Takayoshi},\ and\ \citenamefont {Oka}}]{Sato2016}%
  \BibitemOpen
  \bibfield  {author} {\bibinfo {author} {\bibfnamefont {M.}~\bibnamefont
  {Sato}}, \bibinfo {author} {\bibfnamefont {S.}~\bibnamefont {Takayoshi}}, \
  and\ \bibinfo {author} {\bibfnamefont {T.}~\bibnamefont {Oka}},\ }\href
  {\doibase 10.1103/PhysRevLett.117.147202} {\bibfield  {journal} {\bibinfo
  {journal} {Phys. Rev. Lett.}\ }\textbf {\bibinfo {volume} {117}},\ \bibinfo
  {pages} {147202} (\bibinfo {year} {2016})}\BibitemShut {NoStop}%
\bibitem [{\citenamefont {Kitamura}\ \emph {et~al.}(2017)\citenamefont
  {Kitamura}, \citenamefont {Oka},\ and\ \citenamefont {Aoki}}]{Kitamura2017}%
  \BibitemOpen
  \bibfield  {author} {\bibinfo {author} {\bibfnamefont {S.}~\bibnamefont
  {Kitamura}}, \bibinfo {author} {\bibfnamefont {T.}~\bibnamefont {Oka}}, \
  and\ \bibinfo {author} {\bibfnamefont {H.}~\bibnamefont {Aoki}},\ }\href
  {\doibase 10.1103/PhysRevB.96.014406} {\bibfield  {journal} {\bibinfo
  {journal} {Phys. Rev. B}\ }\textbf {\bibinfo {volume} {96}},\ \bibinfo
  {pages} {014406} (\bibinfo {year} {2017})}\BibitemShut {NoStop}%
\bibitem [{\citenamefont {Claassen}\ \emph {et~al.}(2017)\citenamefont
  {Claassen}, \citenamefont {Jiang}, \citenamefont {Moritz},\ and\
  \citenamefont {Devereaux}}]{Claassen2017}%
  \BibitemOpen
  \bibfield  {author} {\bibinfo {author} {\bibfnamefont {M.}~\bibnamefont
  {Claassen}}, \bibinfo {author} {\bibfnamefont {H.-C.}\ \bibnamefont {Jiang}},
  \bibinfo {author} {\bibfnamefont {B.}~\bibnamefont {Moritz}}, \ and\ \bibinfo
  {author} {\bibfnamefont {T.~P.}\ \bibnamefont {Devereaux}},\ }\href {\doibase
  10.1038/s41467-017-00876-y} {\bibfield  {journal} {\bibinfo  {journal}
  {Nature Communications}\ }\textbf {\bibinfo {volume} {8}},\ \bibinfo {pages}
  {1192} (\bibinfo {year} {2017})}\BibitemShut {NoStop}%
\bibitem [{\citenamefont {{Eckstein}}\ \emph {et~al.}(2017)\citenamefont
  {{Eckstein}}, \citenamefont {{Mentink}},\ and\ \citenamefont
  {{Werner}}}]{Eckstein2017}%
  \BibitemOpen
  \bibfield  {author} {\bibinfo {author} {\bibfnamefont {M.}~\bibnamefont
  {{Eckstein}}}, \bibinfo {author} {\bibfnamefont {J.~H.}\ \bibnamefont
  {{Mentink}}}, \ and\ \bibinfo {author} {\bibfnamefont {P.}~\bibnamefont
  {{Werner}}},\ }\href {https://arxiv.org/abs/1703.03269} {\bibfield  {journal}
  {\bibinfo  {journal} {arXiv : 1703.03269}\ } (\bibinfo {year}
  {2017})}\BibitemShut {NoStop}%
\bibitem [{\citenamefont {G{\"o}rg}\ \emph {et~al.}(2018)\citenamefont
  {G{\"o}rg}, \citenamefont {Messer}, \citenamefont {Sandholzer}, \citenamefont
  {Jotzu}, \citenamefont {Desbuquois},\ and\ \citenamefont
  {Esslinger}}]{Gorg2018}%
  \BibitemOpen
  \bibfield  {author} {\bibinfo {author} {\bibfnamefont {F.}~\bibnamefont
  {G{\"o}rg}}, \bibinfo {author} {\bibfnamefont {M.}~\bibnamefont {Messer}},
  \bibinfo {author} {\bibfnamefont {K.}~\bibnamefont {Sandholzer}}, \bibinfo
  {author} {\bibfnamefont {G.}~\bibnamefont {Jotzu}}, \bibinfo {author}
  {\bibfnamefont {R.}~\bibnamefont {Desbuquois}}, \ and\ \bibinfo {author}
  {\bibfnamefont {T.}~\bibnamefont {Esslinger}},\ }\href
  {http://dx.doi.org/10.1038/nature25135} {\bibfield  {journal} {\bibinfo
  {journal} {Nature}\ }\textbf {\bibinfo {volume} {553}},\ \bibinfo {pages}
  {481 EP } (\bibinfo {year} {2018})}\BibitemShut {NoStop}%
\bibitem [{\citenamefont {Knap}\ \emph {et~al.}(2016)\citenamefont {Knap},
  \citenamefont {Babadi}, \citenamefont {Refael}, \citenamefont {Martin},\ and\
  \citenamefont {Demler}}]{Knap2016}%
  \BibitemOpen
  \bibfield  {author} {\bibinfo {author} {\bibfnamefont {M.}~\bibnamefont
  {Knap}}, \bibinfo {author} {\bibfnamefont {M.}~\bibnamefont {Babadi}},
  \bibinfo {author} {\bibfnamefont {G.}~\bibnamefont {Refael}}, \bibinfo
  {author} {\bibfnamefont {I.}~\bibnamefont {Martin}}, \ and\ \bibinfo {author}
  {\bibfnamefont {E.}~\bibnamefont {Demler}},\ }\href {\doibase
  10.1103/PhysRevB.94.214504} {\bibfield  {journal} {\bibinfo  {journal} {Phys.
  Rev. B}\ }\textbf {\bibinfo {volume} {94}},\ \bibinfo {pages} {214504}
  (\bibinfo {year} {2016})}\BibitemShut {NoStop}%
\bibitem [{\citenamefont {Murakami}\ \emph {et~al.}(2017)\citenamefont
  {Murakami}, \citenamefont {Tsuji}, \citenamefont {Eckstein},\ and\
  \citenamefont {Werner}}]{Murakami2017}%
  \BibitemOpen
  \bibfield  {author} {\bibinfo {author} {\bibfnamefont {Y.}~\bibnamefont
  {Murakami}}, \bibinfo {author} {\bibfnamefont {N.}~\bibnamefont {Tsuji}},
  \bibinfo {author} {\bibfnamefont {M.}~\bibnamefont {Eckstein}}, \ and\
  \bibinfo {author} {\bibfnamefont {P.}~\bibnamefont {Werner}},\ }\href
  {\doibase 10.1103/PhysRevB.96.045125} {\bibfield  {journal} {\bibinfo
  {journal} {Phys. Rev. B}\ }\textbf {\bibinfo {volume} {96}},\ \bibinfo
  {pages} {045125} (\bibinfo {year} {2017})}\BibitemShut {NoStop}%
\bibitem [{\citenamefont {Takasan}\ \emph {et~al.}(2017)\citenamefont
  {Takasan}, \citenamefont {Daido}, \citenamefont {Kawakami},\ and\
  \citenamefont {Yanase}}]{Takasan2017}%
  \BibitemOpen
  \bibfield  {author} {\bibinfo {author} {\bibfnamefont {K.}~\bibnamefont
  {Takasan}}, \bibinfo {author} {\bibfnamefont {A.}~\bibnamefont {Daido}},
  \bibinfo {author} {\bibfnamefont {N.}~\bibnamefont {Kawakami}}, \ and\
  \bibinfo {author} {\bibfnamefont {Y.}~\bibnamefont {Yanase}},\ }\href
  {\doibase 10.1103/PhysRevB.95.134508} {\bibfield  {journal} {\bibinfo
  {journal} {Phys. Rev. B}\ }\textbf {\bibinfo {volume} {95}},\ \bibinfo
  {pages} {134508} (\bibinfo {year} {2017})}\BibitemShut {NoStop}%
\bibitem [{\citenamefont {Fausti}\ \emph {et~al.}(2011)\citenamefont {Fausti},
  \citenamefont {Tobey}, \citenamefont {Dean}, \citenamefont {Kaiser},
  \citenamefont {Dienst}, \citenamefont {Hoffmann}, \citenamefont {Pyon},
  \citenamefont {Takayama}, \citenamefont {Takagi},\ and\ \citenamefont
  {Cavalleri}}]{Fausti2011}%
  \BibitemOpen
  \bibfield  {author} {\bibinfo {author} {\bibfnamefont {D.}~\bibnamefont
  {Fausti}}, \bibinfo {author} {\bibfnamefont {R.~I.}\ \bibnamefont {Tobey}},
  \bibinfo {author} {\bibfnamefont {N.}~\bibnamefont {Dean}}, \bibinfo {author}
  {\bibfnamefont {S.}~\bibnamefont {Kaiser}}, \bibinfo {author} {\bibfnamefont
  {A.}~\bibnamefont {Dienst}}, \bibinfo {author} {\bibfnamefont {M.~C.}\
  \bibnamefont {Hoffmann}}, \bibinfo {author} {\bibfnamefont {S.}~\bibnamefont
  {Pyon}}, \bibinfo {author} {\bibfnamefont {T.}~\bibnamefont {Takayama}},
  \bibinfo {author} {\bibfnamefont {H.}~\bibnamefont {Takagi}}, \ and\ \bibinfo
  {author} {\bibfnamefont {A.}~\bibnamefont {Cavalleri}},\ }\href {\doibase
  10.1126/science.1197294} {\bibfield  {journal} {\bibinfo  {journal}
  {Science}\ }\textbf {\bibinfo {volume} {331}},\ \bibinfo {pages} {189}
  (\bibinfo {year} {2011})}\BibitemShut {NoStop}%
\bibitem [{\citenamefont {Mitrano}\ \emph {et~al.}(2016)\citenamefont
  {Mitrano}, \citenamefont {Cantaluppi}, \citenamefont {Nicoletti},
  \citenamefont {Kaiser}, \citenamefont {Perucchi}, \citenamefont {Lupi},
  \citenamefont {Di~Pietro}, \citenamefont {Pontiroli}, \citenamefont
  {Ricc{\`o}}, \citenamefont {Clark}, \citenamefont {Jaksch},\ and\
  \citenamefont {Cavalleri}}]{Mitrano2016}%
  \BibitemOpen
  \bibfield  {author} {\bibinfo {author} {\bibfnamefont {M.}~\bibnamefont
  {Mitrano}}, \bibinfo {author} {\bibfnamefont {A.}~\bibnamefont {Cantaluppi}},
  \bibinfo {author} {\bibfnamefont {D.}~\bibnamefont {Nicoletti}}, \bibinfo
  {author} {\bibfnamefont {S.}~\bibnamefont {Kaiser}}, \bibinfo {author}
  {\bibfnamefont {A.}~\bibnamefont {Perucchi}}, \bibinfo {author}
  {\bibfnamefont {S.}~\bibnamefont {Lupi}}, \bibinfo {author} {\bibfnamefont
  {P.}~\bibnamefont {Di~Pietro}}, \bibinfo {author} {\bibfnamefont
  {D.}~\bibnamefont {Pontiroli}}, \bibinfo {author} {\bibfnamefont
  {M.}~\bibnamefont {Ricc{\`o}}}, \bibinfo {author} {\bibfnamefont {S.~R.}\
  \bibnamefont {Clark}}, \bibinfo {author} {\bibfnamefont {D.}~\bibnamefont
  {Jaksch}}, \ and\ \bibinfo {author} {\bibfnamefont {A.}~\bibnamefont
  {Cavalleri}},\ }\href {http://dx.doi.org/10.1038/nature16522} {\bibfield
  {journal} {\bibinfo  {journal} {Nature}\ }\textbf {\bibinfo {volume} {530}},\
  \bibinfo {pages} {461 EP } (\bibinfo {year} {2016})}\BibitemShut {NoStop}%
\bibitem [{\citenamefont {Matsunaga}\ \emph {et~al.}(2014)\citenamefont
  {Matsunaga}, \citenamefont {Tsuji}, \citenamefont {Fujita}, \citenamefont
  {Sugioka}, \citenamefont {Makise}, \citenamefont {Uzawa}, \citenamefont
  {Terai}, \citenamefont {Wang}, \citenamefont {Aoki},\ and\ \citenamefont
  {Shimano}}]{Matsunaga2014}%
  \BibitemOpen
  \bibfield  {author} {\bibinfo {author} {\bibfnamefont {R.}~\bibnamefont
  {Matsunaga}}, \bibinfo {author} {\bibfnamefont {N.}~\bibnamefont {Tsuji}},
  \bibinfo {author} {\bibfnamefont {H.}~\bibnamefont {Fujita}}, \bibinfo
  {author} {\bibfnamefont {A.}~\bibnamefont {Sugioka}}, \bibinfo {author}
  {\bibfnamefont {K.}~\bibnamefont {Makise}}, \bibinfo {author} {\bibfnamefont
  {Y.}~\bibnamefont {Uzawa}}, \bibinfo {author} {\bibfnamefont
  {H.}~\bibnamefont {Terai}}, \bibinfo {author} {\bibfnamefont
  {Z.}~\bibnamefont {Wang}}, \bibinfo {author} {\bibfnamefont {H.}~\bibnamefont
  {Aoki}}, \ and\ \bibinfo {author} {\bibfnamefont {R.}~\bibnamefont
  {Shimano}},\ }\href {\doibase 10.1126/science.1254697} {\bibfield  {journal}
  {\bibinfo  {journal} {Science}\ }\textbf {\bibinfo {volume} {345}},\ \bibinfo
  {pages} {1145} (\bibinfo {year} {2014})}\BibitemShut {NoStop}%
\bibitem [{\citenamefont {Kimura}\ \emph {et~al.}(2003)\citenamefont {Kimura},
  \citenamefont {Goto}, \citenamefont {Shintani}, \citenamefont {Ishizaka},
  \citenamefont {Arima},\ and\ \citenamefont {Tokura}}]{Kimura2003}%
  \BibitemOpen
  \bibfield  {author} {\bibinfo {author} {\bibfnamefont {T.}~\bibnamefont
  {Kimura}}, \bibinfo {author} {\bibfnamefont {T.}~\bibnamefont {Goto}},
  \bibinfo {author} {\bibfnamefont {H.}~\bibnamefont {Shintani}}, \bibinfo
  {author} {\bibfnamefont {K.}~\bibnamefont {Ishizaka}}, \bibinfo {author}
  {\bibfnamefont {T.}~\bibnamefont {Arima}}, \ and\ \bibinfo {author}
  {\bibfnamefont {Y.}~\bibnamefont {Tokura}},\ }\href
  {http://dx.doi.org/10.1038/nature02018} {\bibfield  {journal} {\bibinfo
  {journal} {Nature}\ }\textbf {\bibinfo {volume} {426}},\ \bibinfo {pages} {55
  EP } (\bibinfo {year} {2003})}\BibitemShut {NoStop}%
\bibitem [{\citenamefont {Katsura}\ \emph {et~al.}(2005)\citenamefont
  {Katsura}, \citenamefont {Nagaosa},\ and\ \citenamefont
  {Balatsky}}]{Katsura2005}%
  \BibitemOpen
  \bibfield  {author} {\bibinfo {author} {\bibfnamefont {H.}~\bibnamefont
  {Katsura}}, \bibinfo {author} {\bibfnamefont {N.}~\bibnamefont {Nagaosa}}, \
  and\ \bibinfo {author} {\bibfnamefont {A.~V.}\ \bibnamefont {Balatsky}},\
  }\href {\doibase 10.1103/PhysRevLett.95.057205} {\bibfield  {journal}
  {\bibinfo  {journal} {Phys. Rev. Lett.}\ }\textbf {\bibinfo {volume} {95}},\
  \bibinfo {pages} {057205} (\bibinfo {year} {2005})}\BibitemShut {NoStop}%
\bibitem [{\citenamefont {Mostovoy}(2006)}]{Mostovoy2006}%
  \BibitemOpen
  \bibfield  {author} {\bibinfo {author} {\bibfnamefont {M.}~\bibnamefont
  {Mostovoy}},\ }\href {\doibase 10.1103/PhysRevLett.96.067601} {\bibfield
  {journal} {\bibinfo  {journal} {Phys. Rev. Lett.}\ }\textbf {\bibinfo
  {volume} {96}},\ \bibinfo {pages} {067601} (\bibinfo {year}
  {2006})}\BibitemShut {NoStop}%
\bibitem [{\citenamefont {Pimenov}\ \emph {et~al.}(2006)\citenamefont
  {Pimenov}, \citenamefont {Mukhin}, \citenamefont {Ivanov}, \citenamefont
  {Travkin}, \citenamefont {Balbashov},\ and\ \citenamefont
  {Loidl}}]{Pimenov2006}%
  \BibitemOpen
  \bibfield  {author} {\bibinfo {author} {\bibfnamefont {A.}~\bibnamefont
  {Pimenov}}, \bibinfo {author} {\bibfnamefont {A.~A.}\ \bibnamefont {Mukhin}},
  \bibinfo {author} {\bibfnamefont {V.~Y.}\ \bibnamefont {Ivanov}}, \bibinfo
  {author} {\bibfnamefont {V.~D.}\ \bibnamefont {Travkin}}, \bibinfo {author}
  {\bibfnamefont {A.~M.}\ \bibnamefont {Balbashov}}, \ and\ \bibinfo {author}
  {\bibfnamefont {A.}~\bibnamefont {Loidl}},\ }\href
  {http://dx.doi.org/10.1038/nphys212} {\bibfield  {journal} {\bibinfo
  {journal} {Nature Physics}\ }\textbf {\bibinfo {volume} {2}},\ \bibinfo
  {pages} {97 EP } (\bibinfo {year} {2006})}\BibitemShut {NoStop}%
\bibitem [{\citenamefont {Tokunaga}\ \emph {et~al.}(2012)\citenamefont
  {Tokunaga}, \citenamefont {Taguchi}, \citenamefont {Arima},\ and\
  \citenamefont {Tokura}}]{Tokunaga2012}%
  \BibitemOpen
  \bibfield  {author} {\bibinfo {author} {\bibfnamefont {Y.}~\bibnamefont
  {Tokunaga}}, \bibinfo {author} {\bibfnamefont {Y.}~\bibnamefont {Taguchi}},
  \bibinfo {author} {\bibfnamefont {T.-h.}\ \bibnamefont {Arima}}, \ and\
  \bibinfo {author} {\bibfnamefont {Y.}~\bibnamefont {Tokura}},\ }\href
  {http://dx.doi.org/10.1038/nphys2405} {\bibfield  {journal} {\bibinfo
  {journal} {Nature Physics}\ }\textbf {\bibinfo {volume} {8}},\ \bibinfo
  {pages} {838 EP } (\bibinfo {year} {2012})},\ \bibinfo {note}
  {article}\BibitemShut {NoStop}%
\bibitem [{\citenamefont {Tokura}\ \emph {et~al.}(2014)\citenamefont {Tokura},
  \citenamefont {Seki},\ and\ \citenamefont {Nagaosa}}]{Tokura2014}%
  \BibitemOpen
  \bibfield  {author} {\bibinfo {author} {\bibfnamefont {Y.}~\bibnamefont
  {Tokura}}, \bibinfo {author} {\bibfnamefont {S.}~\bibnamefont {Seki}}, \ and\
  \bibinfo {author} {\bibfnamefont {N.}~\bibnamefont {Nagaosa}},\ }\href
  {http://stacks.iop.org/0034-4885/77/i=7/a=076501} {\bibfield  {journal}
  {\bibinfo  {journal} {Reports on Progress in Physics}\ }\textbf {\bibinfo
  {volume} {77}},\ \bibinfo {pages} {076501} (\bibinfo {year}
  {2014})}\BibitemShut {NoStop}%
\bibitem [{\citenamefont {Fukui}\ and\ \citenamefont
  {Kawakami}(1998)}]{Fukui1998}%
  \BibitemOpen
  \bibfield  {author} {\bibinfo {author} {\bibfnamefont {T.}~\bibnamefont
  {Fukui}}\ and\ \bibinfo {author} {\bibfnamefont {N.}~\bibnamefont
  {Kawakami}},\ }\href {\doibase 10.1103/PhysRevB.58.16051} {\bibfield
  {journal} {\bibinfo  {journal} {Phys. Rev. B}\ }\textbf {\bibinfo {volume}
  {58}},\ \bibinfo {pages} {16051} (\bibinfo {year} {1998})}\BibitemShut
  {NoStop}%
\bibitem [{\citenamefont {Oka}\ \emph {et~al.}(2003)\citenamefont {Oka},
  \citenamefont {Arita},\ and\ \citenamefont {Aoki}}]{Oka2003}%
  \BibitemOpen
  \bibfield  {author} {\bibinfo {author} {\bibfnamefont {T.}~\bibnamefont
  {Oka}}, \bibinfo {author} {\bibfnamefont {R.}~\bibnamefont {Arita}}, \ and\
  \bibinfo {author} {\bibfnamefont {H.}~\bibnamefont {Aoki}},\ }\href {\doibase
  10.1103/PhysRevLett.91.066406} {\bibfield  {journal} {\bibinfo  {journal}
  {Phys. Rev. Lett.}\ }\textbf {\bibinfo {volume} {91}},\ \bibinfo {pages}
  {066406} (\bibinfo {year} {2003})}\BibitemShut {NoStop}%
\bibitem [{\citenamefont {Eckstein}\ \emph {et~al.}(2010)\citenamefont
  {Eckstein}, \citenamefont {Oka},\ and\ \citenamefont
  {Werner}}]{Eckstein2010}%
  \BibitemOpen
  \bibfield  {author} {\bibinfo {author} {\bibfnamefont {M.}~\bibnamefont
  {Eckstein}}, \bibinfo {author} {\bibfnamefont {T.}~\bibnamefont {Oka}}, \
  and\ \bibinfo {author} {\bibfnamefont {P.}~\bibnamefont {Werner}},\ }\href
  {\doibase 10.1103/PhysRevLett.105.146404} {\bibfield  {journal} {\bibinfo
  {journal} {Phys. Rev. Lett.}\ }\textbf {\bibinfo {volume} {105}},\ \bibinfo
  {pages} {146404} (\bibinfo {year} {2010})}\BibitemShut {NoStop}%
\bibitem [{\citenamefont {Oka}(2012)}]{Oka2012}%
  \BibitemOpen
  \bibfield  {author} {\bibinfo {author} {\bibfnamefont {T.}~\bibnamefont
  {Oka}},\ }\href {\doibase 10.1103/PhysRevB.86.075148} {\bibfield  {journal}
  {\bibinfo  {journal} {Phys. Rev. B}\ }\textbf {\bibinfo {volume} {86}},\
  \bibinfo {pages} {075148} (\bibinfo {year} {2012})}\BibitemShut {NoStop}%
\bibitem [{\citenamefont {Yamakawa}\ \emph {et~al.}(2017)\citenamefont
  {Yamakawa}, \citenamefont {Miyamoto}, \citenamefont {Morimoto}, \citenamefont
  {Terashige}, \citenamefont {Yada}, \citenamefont {Kida}, \citenamefont
  {Suda}, \citenamefont {Yamamoto}, \citenamefont {Kato}, \citenamefont
  {Miyagawa}, \citenamefont {Kanoda},\ and\ \citenamefont
  {Okamoto}}]{Yamakawa2017}%
  \BibitemOpen
  \bibfield  {author} {\bibinfo {author} {\bibfnamefont {H.}~\bibnamefont
  {Yamakawa}}, \bibinfo {author} {\bibfnamefont {T.}~\bibnamefont {Miyamoto}},
  \bibinfo {author} {\bibfnamefont {T.}~\bibnamefont {Morimoto}}, \bibinfo
  {author} {\bibfnamefont {T.}~\bibnamefont {Terashige}}, \bibinfo {author}
  {\bibfnamefont {H.}~\bibnamefont {Yada}}, \bibinfo {author} {\bibfnamefont
  {N.}~\bibnamefont {Kida}}, \bibinfo {author} {\bibfnamefont {M.}~\bibnamefont
  {Suda}}, \bibinfo {author} {\bibfnamefont {H.~.~M.}\ \bibnamefont
  {Yamamoto}}, \bibinfo {author} {\bibfnamefont {R.}~\bibnamefont {Kato}},
  \bibinfo {author} {\bibfnamefont {K.}~\bibnamefont {Miyagawa}}, \bibinfo
  {author} {\bibfnamefont {K.}~\bibnamefont {Kanoda}}, \ and\ \bibinfo {author}
  {\bibfnamefont {H.}~\bibnamefont {Okamoto}},\ }\href
  {http://dx.doi.org/10.1038/nmat4967} {\bibfield  {journal} {\bibinfo
  {journal} {Nature Materials}\ }\textbf {\bibinfo {volume} {16}},\ \bibinfo
  {pages} {1100 EP } (\bibinfo {year} {2017})}\BibitemShut {NoStop}%
\bibitem [{\citenamefont {Sow}\ \emph {et~al.}(2017)\citenamefont {Sow},
  \citenamefont {Yonezawa}, \citenamefont {Kitamura}, \citenamefont {Oka},
  \citenamefont {Kuroki}, \citenamefont {Nakamura},\ and\ \citenamefont
  {Maeno}}]{Sow2017}%
  \BibitemOpen
  \bibfield  {author} {\bibinfo {author} {\bibfnamefont {C.}~\bibnamefont
  {Sow}}, \bibinfo {author} {\bibfnamefont {S.}~\bibnamefont {Yonezawa}},
  \bibinfo {author} {\bibfnamefont {S.}~\bibnamefont {Kitamura}}, \bibinfo
  {author} {\bibfnamefont {T.}~\bibnamefont {Oka}}, \bibinfo {author}
  {\bibfnamefont {K.}~\bibnamefont {Kuroki}}, \bibinfo {author} {\bibfnamefont
  {F.}~\bibnamefont {Nakamura}}, \ and\ \bibinfo {author} {\bibfnamefont
  {Y.}~\bibnamefont {Maeno}},\ }\href {\doibase 10.1126/science.aah4297}
  {\bibfield  {journal} {\bibinfo  {journal} {Science}\ }\textbf {\bibinfo
  {volume} {358}},\ \bibinfo {pages} {1084} (\bibinfo {year}
  {2017})}\BibitemShut {NoStop}%
\bibitem [{\citenamefont {Ueno}\ \emph {et~al.}(2014)\citenamefont {Ueno},
  \citenamefont {Shimotani}, \citenamefont {Yuan}, \citenamefont {Ye},
  \citenamefont {Kawasaki},\ and\ \citenamefont {Iwasa}}]{Ueno2014}%
  \BibitemOpen
  \bibfield  {author} {\bibinfo {author} {\bibfnamefont {K.}~\bibnamefont
  {Ueno}}, \bibinfo {author} {\bibfnamefont {H.}~\bibnamefont {Shimotani}},
  \bibinfo {author} {\bibfnamefont {H.}~\bibnamefont {Yuan}}, \bibinfo {author}
  {\bibfnamefont {J.}~\bibnamefont {Ye}}, \bibinfo {author} {\bibfnamefont
  {M.}~\bibnamefont {Kawasaki}}, \ and\ \bibinfo {author} {\bibfnamefont
  {Y.}~\bibnamefont {Iwasa}},\ }\href {\doibase 10.7566/JPSJ.83.032001}
  {\bibfield  {journal} {\bibinfo  {journal} {Journal of the Physical Society
  of Japan}\ }\textbf {\bibinfo {volume} {83}},\ \bibinfo {pages} {032001}
  (\bibinfo {year} {2014})}\BibitemShut {NoStop}%
\bibitem [{\citenamefont {Bisri}\ \emph {et~al.}(2017)\citenamefont {Bisri},
  \citenamefont {Shimizu}, \citenamefont {Nakano},\ and\ \citenamefont
  {Iwasa}}]{Bisri2017}%
  \BibitemOpen
  \bibfield  {author} {\bibinfo {author} {\bibfnamefont {S.~Z.}\ \bibnamefont
  {Bisri}}, \bibinfo {author} {\bibfnamefont {S.}~\bibnamefont {Shimizu}},
  \bibinfo {author} {\bibfnamefont {M.}~\bibnamefont {Nakano}}, \ and\ \bibinfo
  {author} {\bibfnamefont {Y.}~\bibnamefont {Iwasa}},\ }\href {\doibase
  10.1002/adma.201607054} {\bibfield  {journal} {\bibinfo  {journal} {Advanced
  Materials}\ }\textbf {\bibinfo {volume} {29}},\ \bibinfo {pages} {1607054}
  (\bibinfo {year} {2017})},\ \bibinfo {note} {1607054}\BibitemShut {NoStop}%
\bibitem [{\citenamefont {Hsu}\ \emph {et~al.}(2016)\citenamefont {Hsu},
  \citenamefont {Kubetzka}, \citenamefont {Finco}, \citenamefont {Romming},
  \citenamefont {von Bergmann},\ and\ \citenamefont {Wiesendanger}}]{Hsu2016}%
  \BibitemOpen
  \bibfield  {author} {\bibinfo {author} {\bibfnamefont {P.-J.}\ \bibnamefont
  {Hsu}}, \bibinfo {author} {\bibfnamefont {A.}~\bibnamefont {Kubetzka}},
  \bibinfo {author} {\bibfnamefont {A.}~\bibnamefont {Finco}}, \bibinfo
  {author} {\bibfnamefont {N.}~\bibnamefont {Romming}}, \bibinfo {author}
  {\bibfnamefont {K.}~\bibnamefont {von Bergmann}}, \ and\ \bibinfo {author}
  {\bibfnamefont {R.}~\bibnamefont {Wiesendanger}},\ }\href
  {http://dx.doi.org/10.1038/nnano.2016.234} {\bibfield  {journal} {\bibinfo
  {journal} {Nature Nanotechnology}\ }\textbf {\bibinfo {volume} {12}},\
  \bibinfo {pages} {123 EP } (\bibinfo {year} {2016})}\BibitemShut {NoStop}%
\bibitem [{\citenamefont {Hirori}\ \emph {et~al.}(2011)\citenamefont {Hirori},
  \citenamefont {Doi}, \citenamefont {Blanchard},\ and\ \citenamefont
  {Tanaka}}]{Hirori2011}%
  \BibitemOpen
  \bibfield  {author} {\bibinfo {author} {\bibfnamefont {H.}~\bibnamefont
  {Hirori}}, \bibinfo {author} {\bibfnamefont {A.}~\bibnamefont {Doi}},
  \bibinfo {author} {\bibfnamefont {F.}~\bibnamefont {Blanchard}}, \ and\
  \bibinfo {author} {\bibfnamefont {K.}~\bibnamefont {Tanaka}},\ }\href
  {\doibase 10.1063/1.3560062} {\bibfield  {journal} {\bibinfo  {journal}
  {Applied Physics Letters}\ }\textbf {\bibinfo {volume} {98}},\ \bibinfo
  {pages} {091106} (\bibinfo {year} {2011})}\BibitemShut {NoStop}%
\bibitem [{\citenamefont {Nicoletti}\ and\ \citenamefont
  {Cavalleri}(2016)}]{Nicoletti2016}%
  \BibitemOpen
  \bibfield  {author} {\bibinfo {author} {\bibfnamefont {D.}~\bibnamefont
  {Nicoletti}}\ and\ \bibinfo {author} {\bibfnamefont {A.}~\bibnamefont
  {Cavalleri}},\ }\href {\doibase 10.1364/AOP.8.000401} {\bibfield  {journal}
  {\bibinfo  {journal} {Adv. Opt. Photon.}\ }\textbf {\bibinfo {volume} {8}},\
  \bibinfo {pages} {401} (\bibinfo {year} {2016})}\BibitemShut {NoStop}%
\bibitem [{\citenamefont {Savary}\ and\ \citenamefont
  {Balents}(2017)}]{Savary2017}%
  \BibitemOpen
  \bibfield  {author} {\bibinfo {author} {\bibfnamefont {L.}~\bibnamefont
  {Savary}}\ and\ \bibinfo {author} {\bibfnamefont {L.}~\bibnamefont
  {Balents}},\ }\href {http://stacks.iop.org/0034-4885/80/i=1/a=016502}
  {\bibfield  {journal} {\bibinfo  {journal} {Reports on Progress in Physics}\
  }\textbf {\bibinfo {volume} {80}},\ \bibinfo {pages} {016502} (\bibinfo
  {year} {2017})}\BibitemShut {NoStop}%
\bibitem [{\citenamefont {Zhou}\ \emph {et~al.}(2017)\citenamefont {Zhou},
  \citenamefont {Kanoda},\ and\ \citenamefont {Ng}}]{Zhou2017}%
  \BibitemOpen
  \bibfield  {author} {\bibinfo {author} {\bibfnamefont {Y.}~\bibnamefont
  {Zhou}}, \bibinfo {author} {\bibfnamefont {K.}~\bibnamefont {Kanoda}}, \ and\
  \bibinfo {author} {\bibfnamefont {T.-K.}\ \bibnamefont {Ng}},\ }\href
  {\doibase 10.1103/RevModPhys.89.025003} {\bibfield  {journal} {\bibinfo
  {journal} {Rev. Mod. Phys.}\ }\textbf {\bibinfo {volume} {89}},\ \bibinfo
  {pages} {025003} (\bibinfo {year} {2017})}\BibitemShut {NoStop}%
\bibitem [{\citenamefont {Wen}(2007)}]{XGWen_book}%
  \BibitemOpen
  \bibfield  {author} {\bibinfo {author} {\bibfnamefont {X.~G.}\ \bibnamefont
  {Wen}},\ }\href@noop {} {\emph {\bibinfo {title} {{Quantum Field Theory of
  Many-body Systems: From the Origin of Sound to an Origin of Light and
  Electrons}}}}\ (\bibinfo  {publisher} {Oxford University Press, New York},\
  \bibinfo {year} {2007})\BibitemShut {NoStop}%
\bibitem [{\citenamefont {Haldane}(1983{\natexlab{a}})}]{Haldane1983a}%
  \BibitemOpen
  \bibfield  {author} {\bibinfo {author} {\bibfnamefont {F.~D.~M.}\
  \bibnamefont {Haldane}},\ }\href {\doibase 10.1103/PhysRevLett.50.1153}
  {\bibfield  {journal} {\bibinfo  {journal} {Phys. Rev. Lett.}\ }\textbf
  {\bibinfo {volume} {50}},\ \bibinfo {pages} {1153} (\bibinfo {year}
  {1983}{\natexlab{a}})}\BibitemShut {NoStop}%
\bibitem [{\citenamefont {Haldane}(1983{\natexlab{b}})}]{Haldane1983b}%
  \BibitemOpen
  \bibfield  {author} {\bibinfo {author} {\bibfnamefont {F.}~\bibnamefont
  {Haldane}},\ }\href {\doibase https://doi.org/10.1016/0375-9601(83)90631-X}
  {\bibfield  {journal} {\bibinfo  {journal} {Physics Letters A}\ }\textbf
  {\bibinfo {volume} {93}},\ \bibinfo {pages} {464 } (\bibinfo {year}
  {1983}{\natexlab{b}})}\BibitemShut {NoStop}%
\bibitem [{\citenamefont {Affleck}\ \emph {et~al.}(1987)\citenamefont
  {Affleck}, \citenamefont {Kennedy}, \citenamefont {Lieb},\ and\ \citenamefont
  {Tasaki}}]{Affleck1987}%
  \BibitemOpen
  \bibfield  {author} {\bibinfo {author} {\bibfnamefont {I.}~\bibnamefont
  {Affleck}}, \bibinfo {author} {\bibfnamefont {T.}~\bibnamefont {Kennedy}},
  \bibinfo {author} {\bibfnamefont {E.~H.}\ \bibnamefont {Lieb}}, \ and\
  \bibinfo {author} {\bibfnamefont {H.}~\bibnamefont {Tasaki}},\ }\href
  {\doibase 10.1103/PhysRevLett.59.799} {\bibfield  {journal} {\bibinfo
  {journal} {Phys. Rev. Lett.}\ }\textbf {\bibinfo {volume} {59}},\ \bibinfo
  {pages} {799} (\bibinfo {year} {1987})}\BibitemShut {NoStop}%
\bibitem [{\citenamefont {Affleck}(1989)}]{Affleck_Lecture}%
  \BibitemOpen
  \bibfield  {author} {\bibinfo {author} {\bibfnamefont {I.}~\bibnamefont
  {Affleck}},\ }\href@noop {} {\emph {\bibinfo {title} {{in Champs, Cordes et
  Phenomenes Critiques; Fields, Strings and Critical Phenomena} edited by E.
  Brézin and J. Zinn-Justin}}}\ (\bibinfo  {publisher} {Elsevier, Amsterdam},\
  \bibinfo {year} {1989})\ p.\ \bibinfo {pages} {564}\BibitemShut {NoStop}%
\bibitem [{\citenamefont {Pollmann}\ \emph {et~al.}(2010)\citenamefont
  {Pollmann}, \citenamefont {Turner}, \citenamefont {Berg},\ and\ \citenamefont
  {Oshikawa}}]{Pollmann2010}%
  \BibitemOpen
  \bibfield  {author} {\bibinfo {author} {\bibfnamefont {F.}~\bibnamefont
  {Pollmann}}, \bibinfo {author} {\bibfnamefont {A.~M.}\ \bibnamefont
  {Turner}}, \bibinfo {author} {\bibfnamefont {E.}~\bibnamefont {Berg}}, \ and\
  \bibinfo {author} {\bibfnamefont {M.}~\bibnamefont {Oshikawa}},\ }\href
  {\doibase 10.1103/PhysRevB.81.064439} {\bibfield  {journal} {\bibinfo
  {journal} {Phys. Rev. B}\ }\textbf {\bibinfo {volume} {81}},\ \bibinfo
  {pages} {064439} (\bibinfo {year} {2010})}\BibitemShut {NoStop}%
\bibitem [{\citenamefont {Pollmann}\ \emph {et~al.}(2012)\citenamefont
  {Pollmann}, \citenamefont {Berg}, \citenamefont {Turner},\ and\ \citenamefont
  {Oshikawa}}]{Pollmann2012}%
  \BibitemOpen
  \bibfield  {author} {\bibinfo {author} {\bibfnamefont {F.}~\bibnamefont
  {Pollmann}}, \bibinfo {author} {\bibfnamefont {E.}~\bibnamefont {Berg}},
  \bibinfo {author} {\bibfnamefont {A.~M.}\ \bibnamefont {Turner}}, \ and\
  \bibinfo {author} {\bibfnamefont {M.}~\bibnamefont {Oshikawa}},\ }\href
  {\doibase 10.1103/PhysRevB.85.075125} {\bibfield  {journal} {\bibinfo
  {journal} {Phys. Rev. B}\ }\textbf {\bibinfo {volume} {85}},\ \bibinfo
  {pages} {075125} (\bibinfo {year} {2012})}\BibitemShut {NoStop}%
\bibitem [{\citenamefont {Giamarchi}(2003)}]{Giamarchi_book}%
  \BibitemOpen
  \bibfield  {author} {\bibinfo {author} {\bibfnamefont {T.}~\bibnamefont
  {Giamarchi}},\ }\href@noop {} {\emph {\bibinfo {title} {Quantum Physics in
  One Dimension}}}\ (\bibinfo  {publisher} {Oxford University Press, New
  York},\ \bibinfo {year} {2003})\BibitemShut {NoStop}%
\bibitem [{\citenamefont {Gogolin}\ \emph {et~al.}(2004)\citenamefont
  {Gogolin}, \citenamefont {Nersesian},\ and\ \citenamefont
  {Tsvelik}}]{Gogolin_book}%
  \BibitemOpen
  \bibfield  {author} {\bibinfo {author} {\bibfnamefont {A.~O.}\ \bibnamefont
  {Gogolin}}, \bibinfo {author} {\bibfnamefont {A.~A.}\ \bibnamefont
  {Nersesian}}, \ and\ \bibinfo {author} {\bibfnamefont {A.~M.}\ \bibnamefont
  {Tsvelik}},\ }\href@noop {} {\emph {\bibinfo {title} {{Bosonization and
  strongly correlated systems}}}}\ (\bibinfo  {publisher} {Cambridge University
  Press, UK},\ \bibinfo {year} {2004})\BibitemShut {NoStop}%
\bibitem [{\citenamefont {Tsvelik}(2007)}]{Tsvelik_book}%
  \BibitemOpen
  \bibfield  {author} {\bibinfo {author} {\bibfnamefont {A.~M.}\ \bibnamefont
  {Tsvelik}},\ }\href@noop {} {\emph {\bibinfo {title} {{Quantum Field Theory
  in Condensed Matter Physics}}}}\ (\bibinfo  {publisher} {Cambridge University
  Press, UK},\ \bibinfo {year} {2007})\BibitemShut {NoStop}%
\bibitem{FN1}
  Strictly speaking, we should take into account the dielectric constant $\epsilon$ defined for each material. In other words, the electric field $\bm E$ used here is $D=\epsilon E$.
\bibitem{FN2}
  The detail of the derivation for the spin-$\frac{1}{2}$ model is given in Supplemental Material.
\bibitem{FN3}
  DC-field driven modifications of exchange couplings have been studied in a few specific systems. See, for example, H. Katsura, M. Sato, T. Furuta, and N. Nagaosa, Phys. Rev. Lett. \textbf{103}, 177402 (2009). However, we stress that our present results could be applied in a generic class of Mott insulators.
\bibitem [{\citenamefont {Bloch}\ \emph {et~al.}(2008)\citenamefont {Bloch},
  \citenamefont {Dalibard},\ and\ \citenamefont {Zwerger}}]{Bloch2008}%
  \BibitemOpen
  \bibfield  {author} {\bibinfo {author} {\bibfnamefont {I.}~\bibnamefont
  {Bloch}}, \bibinfo {author} {\bibfnamefont {J.}~\bibnamefont {Dalibard}}, \
  and\ \bibinfo {author} {\bibfnamefont {W.}~\bibnamefont {Zwerger}},\ }\href
  {\doibase 10.1103/RevModPhys.80.885} {\bibfield  {journal} {\bibinfo
  {journal} {Rev. Mod. Phys.}\ }\textbf {\bibinfo {volume} {80}},\ \bibinfo
  {pages} {885} (\bibinfo {year} {2008})}\BibitemShut {NoStop}%
\bibitem [{\citenamefont {Greiner}\ \emph {et~al.}(2002)\citenamefont
  {Greiner}, \citenamefont {Mandel}, \citenamefont {Esslinger}, \citenamefont
  {H{\"a}nsch},\ and\ \citenamefont {Bloch}}]{Greiner2002}%
  \BibitemOpen
  \bibfield  {author} {\bibinfo {author} {\bibfnamefont {M.}~\bibnamefont
  {Greiner}}, \bibinfo {author} {\bibfnamefont {O.}~\bibnamefont {Mandel}},
  \bibinfo {author} {\bibfnamefont {T.}~\bibnamefont {Esslinger}}, \bibinfo
  {author} {\bibfnamefont {T.~W.}\ \bibnamefont {H{\"a}nsch}}, \ and\ \bibinfo
  {author} {\bibfnamefont {I.}~\bibnamefont {Bloch}},\ }\href
  {http://dx.doi.org/10.1038/415039a} {\bibfield  {journal} {\bibinfo
  {journal} {Nature}\ }\textbf {\bibinfo {volume} {415}},\ \bibinfo {pages} {39
  EP } (\bibinfo {year} {2002})},\ \bibinfo {note} {article}\BibitemShut
  {NoStop}%
\bibitem [{\citenamefont {Mazurenko}\ \emph {et~al.}(2017)\citenamefont
  {Mazurenko}, \citenamefont {Chiu}, \citenamefont {Ji}, \citenamefont
  {Parsons}, \citenamefont {Kan{\'a}sz-Nagy}, \citenamefont {Schmidt},
  \citenamefont {Grusdt}, \citenamefont {Demler}, \citenamefont {Greif},\ and\
  \citenamefont {Greiner}}]{Mazurenko2017}%
  \BibitemOpen
  \bibfield  {author} {\bibinfo {author} {\bibfnamefont {A.}~\bibnamefont
  {Mazurenko}}, \bibinfo {author} {\bibfnamefont {C.~S.}\ \bibnamefont {Chiu}},
  \bibinfo {author} {\bibfnamefont {G.}~\bibnamefont {Ji}}, \bibinfo {author}
  {\bibfnamefont {M.~F.}\ \bibnamefont {Parsons}}, \bibinfo {author}
  {\bibfnamefont {M.}~\bibnamefont {Kan{\'a}sz-Nagy}}, \bibinfo {author}
  {\bibfnamefont {R.}~\bibnamefont {Schmidt}}, \bibinfo {author} {\bibfnamefont
  {F.}~\bibnamefont {Grusdt}}, \bibinfo {author} {\bibfnamefont
  {E.}~\bibnamefont {Demler}}, \bibinfo {author} {\bibfnamefont
  {D.}~\bibnamefont {Greif}}, \ and\ \bibinfo {author} {\bibfnamefont
  {M.}~\bibnamefont {Greiner}},\ }\href {http://dx.doi.org/10.1038/nature22362}
  {\bibfield  {journal} {\bibinfo  {journal} {Nature}\ }\textbf {\bibinfo
  {volume} {545}},\ \bibinfo {pages} {462 EP } (\bibinfo {year}
  {2017})}\BibitemShut {NoStop}%
\bibitem{FN4}
  The detail of the derivation for the spin-1 model is also given in Supplemental Material.
\bibitem{FN5}
  The control of the exchange coupling have been discussed with AC electric fields \cite{Mentink2015, Kitamura2017, Eckstein2017, Claassen2017, Gorg2018}. One of our results (\ref{eq:spin-1/2}) is consistent with the DC limit ($\omega \rightarrow 0$) in the preceding study~\cite{Eckstein2017}. We however stress that our method can be straightforwardly extended to a generic class of Mott insulators. 
\bibitem [{\citenamefont {Nitta}\ \emph {et~al.}(1997)\citenamefont {Nitta},
  \citenamefont {Akazaki}, \citenamefont {Takayanagi},\ and\ \citenamefont
  {Enoki}}]{Nitta1997}%
  \BibitemOpen
  \bibfield  {author} {\bibinfo {author} {\bibfnamefont {J.}~\bibnamefont
  {Nitta}}, \bibinfo {author} {\bibfnamefont {T.}~\bibnamefont {Akazaki}},
  \bibinfo {author} {\bibfnamefont {H.}~\bibnamefont {Takayanagi}}, \ and\
  \bibinfo {author} {\bibfnamefont {T.}~\bibnamefont {Enoki}},\ }\href
  {\doibase 10.1103/PhysRevLett.78.1335} {\bibfield  {journal} {\bibinfo
  {journal} {Phys. Rev. Lett.}\ }\textbf {\bibinfo {volume} {78}},\ \bibinfo
  {pages} {1335} (\bibinfo {year} {1997})}\BibitemShut {NoStop}%
\bibitem [{\citenamefont {Caviglia}\ \emph {et~al.}(2010)\citenamefont
  {Caviglia}, \citenamefont {Gabay}, \citenamefont {Gariglio}, \citenamefont
  {Reyren}, \citenamefont {Cancellieri},\ and\ \citenamefont
  {Triscone}}]{Caviglia2010}%
  \BibitemOpen
  \bibfield  {author} {\bibinfo {author} {\bibfnamefont {A.~D.}\ \bibnamefont
  {Caviglia}}, \bibinfo {author} {\bibfnamefont {M.}~\bibnamefont {Gabay}},
  \bibinfo {author} {\bibfnamefont {S.}~\bibnamefont {Gariglio}}, \bibinfo
  {author} {\bibfnamefont {N.}~\bibnamefont {Reyren}}, \bibinfo {author}
  {\bibfnamefont {C.}~\bibnamefont {Cancellieri}}, \ and\ \bibinfo {author}
  {\bibfnamefont {J.-M.}\ \bibnamefont {Triscone}},\ }\href {\doibase
  10.1103/PhysRevLett.104.126803} {\bibfield  {journal} {\bibinfo  {journal}
  {Phys. Rev. Lett.}\ }\textbf {\bibinfo {volume} {104}},\ \bibinfo {pages}
  {126803} (\bibinfo {year} {2010})}\BibitemShut {NoStop}%
\bibitem [{\citenamefont {Soumyanarayanan}\ \emph {et~al.}(2016)\citenamefont
  {Soumyanarayanan}, \citenamefont {Reyren}, \citenamefont {Fert},\ and\
  \citenamefont {Panagopoulos}}]{Soumyanarayanan2016}%
  \BibitemOpen
  \bibfield  {author} {\bibinfo {author} {\bibfnamefont {A.}~\bibnamefont
  {Soumyanarayanan}}, \bibinfo {author} {\bibfnamefont {N.}~\bibnamefont
  {Reyren}}, \bibinfo {author} {\bibfnamefont {A.}~\bibnamefont {Fert}}, \ and\
  \bibinfo {author} {\bibfnamefont {C.}~\bibnamefont {Panagopoulos}},\ }\href
  {http://dx.doi.org/10.1038/nature19820} {\bibfield  {journal} {\bibinfo
  {journal} {Nature}\ }\textbf {\bibinfo {volume} {539}},\ \bibinfo {pages}
  {509 EP } (\bibinfo {year} {2016})},\ \bibinfo {note} {review
  Article}\BibitemShut {NoStop}%
\bibitem [{\citenamefont {Jolicoeur}\ and\ \citenamefont
  {Le~Guillou}(1989)}]{Jolicoeur1989}%
  \BibitemOpen
  \bibfield  {author} {\bibinfo {author} {\bibfnamefont {T.}~\bibnamefont
  {Jolicoeur}}\ and\ \bibinfo {author} {\bibfnamefont {J.~C.}\ \bibnamefont
  {Le~Guillou}},\ }\href {\doibase 10.1103/PhysRevB.40.2727} {\bibfield
  {journal} {\bibinfo  {journal} {Phys. Rev. B}\ }\textbf {\bibinfo {volume}
  {40}},\ \bibinfo {pages} {2727} (\bibinfo {year} {1989})}\BibitemShut
  {NoStop}%
\bibitem [{\citenamefont {Neuberger}\ and\ \citenamefont
  {Ziman}(1989)}]{Neuberger1989}%
  \BibitemOpen
  \bibfield  {author} {\bibinfo {author} {\bibfnamefont {H.}~\bibnamefont
  {Neuberger}}\ and\ \bibinfo {author} {\bibfnamefont {T.}~\bibnamefont
  {Ziman}},\ }\href {\doibase 10.1103/PhysRevB.39.2608} {\bibfield  {journal}
  {\bibinfo  {journal} {Phys. Rev. B}\ }\textbf {\bibinfo {volume} {39}},\
  \bibinfo {pages} {2608} (\bibinfo {year} {1989})}\BibitemShut {NoStop}%
\bibitem [{\citenamefont {Bernu}\ \emph {et~al.}(1994)\citenamefont {Bernu},
  \citenamefont {Lecheminant}, \citenamefont {Lhuillier},\ and\ \citenamefont
  {Pierre}}]{Bernu1994}%
  \BibitemOpen
  \bibfield  {author} {\bibinfo {author} {\bibfnamefont {B.}~\bibnamefont
  {Bernu}}, \bibinfo {author} {\bibfnamefont {P.}~\bibnamefont {Lecheminant}},
  \bibinfo {author} {\bibfnamefont {C.}~\bibnamefont {Lhuillier}}, \ and\
  \bibinfo {author} {\bibfnamefont {L.}~\bibnamefont {Pierre}},\ }\href
  {\doibase 10.1103/PhysRevB.50.10048} {\bibfield  {journal} {\bibinfo
  {journal} {Phys. Rev. B}\ }\textbf {\bibinfo {volume} {50}},\ \bibinfo
  {pages} {10048} (\bibinfo {year} {1994})}\BibitemShut {NoStop}%
\bibitem [{\citenamefont {Weihong}\ \emph {et~al.}(1999)\citenamefont
  {Weihong}, \citenamefont {McKenzie},\ and\ \citenamefont
  {Singh}}]{Weihong1999}%
  \BibitemOpen
  \bibfield  {author} {\bibinfo {author} {\bibfnamefont {Z.}~\bibnamefont
  {Weihong}}, \bibinfo {author} {\bibfnamefont {R.~H.}\ \bibnamefont
  {McKenzie}}, \ and\ \bibinfo {author} {\bibfnamefont {R.~R.~P.}\ \bibnamefont
  {Singh}},\ }\href {\doibase 10.1103/PhysRevB.59.14367} {\bibfield  {journal}
  {\bibinfo  {journal} {Phys. Rev. B}\ }\textbf {\bibinfo {volume} {59}},\
  \bibinfo {pages} {14367} (\bibinfo {year} {1999})}\BibitemShut {NoStop}%
\bibitem [{\citenamefont {Yunoki}\ and\ \citenamefont
  {Sorella}(2006)}]{Yunoki2006}%
  \BibitemOpen
  \bibfield  {author} {\bibinfo {author} {\bibfnamefont {S.}~\bibnamefont
  {Yunoki}}\ and\ \bibinfo {author} {\bibfnamefont {S.}~\bibnamefont
  {Sorella}},\ }\href {\doibase 10.1103/PhysRevB.74.014408} {\bibfield
  {journal} {\bibinfo  {journal} {Phys. Rev. B}\ }\textbf {\bibinfo {volume}
  {74}},\ \bibinfo {pages} {014408} (\bibinfo {year} {2006})}\BibitemShut
  {NoStop}%
\bibitem [{\citenamefont {Weng}\ \emph {et~al.}(2006)\citenamefont {Weng},
  \citenamefont {Sheng}, \citenamefont {Weng},\ and\ \citenamefont
  {Bursill}}]{Weng2006}%
  \BibitemOpen
  \bibfield  {author} {\bibinfo {author} {\bibfnamefont {M.~Q.}\ \bibnamefont
  {Weng}}, \bibinfo {author} {\bibfnamefont {D.~N.}\ \bibnamefont {Sheng}},
  \bibinfo {author} {\bibfnamefont {Z.~Y.}\ \bibnamefont {Weng}}, \ and\
  \bibinfo {author} {\bibfnamefont {R.~J.}\ \bibnamefont {Bursill}},\ }\href
  {\doibase 10.1103/PhysRevB.74.012407} {\bibfield  {journal} {\bibinfo
  {journal} {Phys. Rev. B}\ }\textbf {\bibinfo {volume} {74}},\ \bibinfo
  {pages} {012407} (\bibinfo {year} {2006})}\BibitemShut {NoStop}%
\bibitem [{\citenamefont {Heidarian}\ \emph {et~al.}(2009)\citenamefont
  {Heidarian}, \citenamefont {Sorella},\ and\ \citenamefont
  {Becca}}]{Heidarian2009}%
  \BibitemOpen
  \bibfield  {author} {\bibinfo {author} {\bibfnamefont {D.}~\bibnamefont
  {Heidarian}}, \bibinfo {author} {\bibfnamefont {S.}~\bibnamefont {Sorella}},
  \ and\ \bibinfo {author} {\bibfnamefont {F.}~\bibnamefont {Becca}},\ }\href
  {\doibase 10.1103/PhysRevB.80.012404} {\bibfield  {journal} {\bibinfo
  {journal} {Phys. Rev. B}\ }\textbf {\bibinfo {volume} {80}},\ \bibinfo
  {pages} {012404} (\bibinfo {year} {2009})}\BibitemShut {NoStop}%
\bibitem [{\citenamefont {Harada}(2012)}]{Harada2012}%
  \BibitemOpen
  \bibfield  {author} {\bibinfo {author} {\bibfnamefont {K.}~\bibnamefont
  {Harada}},\ }\href {\doibase 10.1103/PhysRevB.86.184421} {\bibfield
  {journal} {\bibinfo  {journal} {Phys. Rev. B}\ }\textbf {\bibinfo {volume}
  {86}},\ \bibinfo {pages} {184421} (\bibinfo {year} {2012})}\BibitemShut
  {NoStop}%
\bibitem [{\citenamefont {Starykh}\ and\ \citenamefont
  {Balents}(2007)}]{Starykh2007}%
  \BibitemOpen
  \bibfield  {author} {\bibinfo {author} {\bibfnamefont {O.~A.}\ \bibnamefont
  {Starykh}}\ and\ \bibinfo {author} {\bibfnamefont {L.}~\bibnamefont
  {Balents}},\ }\href {\doibase 10.1103/PhysRevLett.98.077205} {\bibfield
  {journal} {\bibinfo  {journal} {Phys. Rev. Lett.}\ }\textbf {\bibinfo
  {volume} {98}},\ \bibinfo {pages} {077205} (\bibinfo {year}
  {2007})}\BibitemShut {NoStop}%
\bibitem [{\citenamefont {Schultz}\ \emph {et~al.}(1991)\citenamefont
  {Schultz}, \citenamefont {Beno}, \citenamefont {Geiser}, \citenamefont
  {Wang}, \citenamefont {Kini}, \citenamefont {Williams},\ and\ \citenamefont
  {Whangbo}}]{Schultz1991}%
  \BibitemOpen
  \bibfield  {author} {\bibinfo {author} {\bibfnamefont {A.~J.}\ \bibnamefont
  {Schultz}}, \bibinfo {author} {\bibfnamefont {M.~A.}\ \bibnamefont {Beno}},
  \bibinfo {author} {\bibfnamefont {U.}~\bibnamefont {Geiser}}, \bibinfo
  {author} {\bibfnamefont {H.}~\bibnamefont {Wang}}, \bibinfo {author}
  {\bibfnamefont {A.~M.}\ \bibnamefont {Kini}}, \bibinfo {author}
  {\bibfnamefont {J.~M.}\ \bibnamefont {Williams}}, \ and\ \bibinfo {author}
  {\bibfnamefont {M.-H.}\ \bibnamefont {Whangbo}},\ }\href {\doibase
  https://doi.org/10.1016/0022-4596(91)90201-R} {\bibfield  {journal} {\bibinfo
   {journal} {Journal of Solid State Chemistry}\ }\textbf {\bibinfo {volume}
  {94}},\ \bibinfo {pages} {352 } (\bibinfo {year} {1991})}\BibitemShut
  {NoStop}%
\bibitem [{\citenamefont {Nakamura}\ \emph {et~al.}(2009)\citenamefont
  {Nakamura}, \citenamefont {Yoshimoto}, \citenamefont {Kosugi}, \citenamefont
  {Arita},\ and\ \citenamefont {Imada}}]{Nakamura2009}%
  \BibitemOpen
  \bibfield  {author} {\bibinfo {author} {\bibfnamefont {K.}~\bibnamefont
  {Nakamura}}, \bibinfo {author} {\bibfnamefont {Y.}~\bibnamefont {Yoshimoto}},
  \bibinfo {author} {\bibfnamefont {T.}~\bibnamefont {Kosugi}}, \bibinfo
  {author} {\bibfnamefont {R.}~\bibnamefont {Arita}}, \ and\ \bibinfo {author}
  {\bibfnamefont {M.}~\bibnamefont {Imada}},\ }\href {\doibase
  10.1143/JPSJ.78.083710} {\bibfield  {journal} {\bibinfo  {journal} {Journal
  of the Physical Society of Japan}\ }\textbf {\bibinfo {volume} {78}},\
  \bibinfo {pages} {083710} (\bibinfo {year} {2009})}\BibitemShut {NoStop}%
\bibitem [{\citenamefont {Shimizu}\ \emph {et~al.}(2003)\citenamefont
  {Shimizu}, \citenamefont {Miyagawa}, \citenamefont {Kanoda}, \citenamefont
  {Maesato},\ and\ \citenamefont {Saito}}]{Shimizu2003}%
  \BibitemOpen
  \bibfield  {author} {\bibinfo {author} {\bibfnamefont {Y.}~\bibnamefont
  {Shimizu}}, \bibinfo {author} {\bibfnamefont {K.}~\bibnamefont {Miyagawa}},
  \bibinfo {author} {\bibfnamefont {K.}~\bibnamefont {Kanoda}}, \bibinfo
  {author} {\bibfnamefont {M.}~\bibnamefont {Maesato}}, \ and\ \bibinfo
  {author} {\bibfnamefont {G.}~\bibnamefont {Saito}},\ }\href {\doibase
  10.1103/PhysRevLett.91.107001} {\bibfield  {journal} {\bibinfo  {journal}
  {Phys. Rev. Lett.}\ }\textbf {\bibinfo {volume} {91}},\ \bibinfo {pages}
  {107001} (\bibinfo {year} {2003})}\BibitemShut {NoStop}%
\bibitem [{\citenamefont {K\'ezsm\'arki}\ \emph {et~al.}(2006)\citenamefont
  {K\'ezsm\'arki}, \citenamefont {Shimizu}, \citenamefont {Mih\'aly},
  \citenamefont {Tokura}, \citenamefont {Kanoda},\ and\ \citenamefont
  {Saito}}]{Kezsmarki2006}%
  \BibitemOpen
  \bibfield  {author} {\bibinfo {author} {\bibfnamefont {I.}~\bibnamefont
  {K\'ezsm\'arki}}, \bibinfo {author} {\bibfnamefont {Y.}~\bibnamefont
  {Shimizu}}, \bibinfo {author} {\bibfnamefont {G.}~\bibnamefont {Mih\'aly}},
  \bibinfo {author} {\bibfnamefont {Y.}~\bibnamefont {Tokura}}, \bibinfo
  {author} {\bibfnamefont {K.}~\bibnamefont {Kanoda}}, \ and\ \bibinfo {author}
  {\bibfnamefont {G.}~\bibnamefont {Saito}},\ }\href {\doibase
  10.1103/PhysRevB.74.201101} {\bibfield  {journal} {\bibinfo  {journal} {Phys.
  Rev. B}\ }\textbf {\bibinfo {volume} {74}},\ \bibinfo {pages} {201101}
  (\bibinfo {year} {2006})}\BibitemShut {NoStop}%
\bibitem [{\citenamefont {Mannhart}\ \emph {et~al.}(1993)\citenamefont
  {Mannhart}, \citenamefont {Ströbel}, \citenamefont {Bednorz},\ and\
  \citenamefont {Gerber}}]{Mannhart1993}%
  \BibitemOpen
  \bibfield  {author} {\bibinfo {author} {\bibfnamefont {J.}~\bibnamefont
  {Mannhart}}, \bibinfo {author} {\bibfnamefont {J.}~\bibnamefont {Ströbel}},
  \bibinfo {author} {\bibfnamefont {J.~G.}\ \bibnamefont {Bednorz}}, \ and\
  \bibinfo {author} {\bibfnamefont {C.}~\bibnamefont {Gerber}},\ }\href
  {\doibase 10.1063/1.108877} {\bibfield  {journal} {\bibinfo  {journal}
  {Applied Physics Letters}\ }\textbf {\bibinfo {volume} {62}},\ \bibinfo
  {pages} {630} (\bibinfo {year} {1993})}\BibitemShut {NoStop}%
\bibitem [{\citenamefont {Starykh}\ and\ \citenamefont
  {Balents}(2004)}]{Starykh2004}%
  \BibitemOpen
  \bibfield  {author} {\bibinfo {author} {\bibfnamefont {O.~A.}\ \bibnamefont
  {Starykh}}\ and\ \bibinfo {author} {\bibfnamefont {L.}~\bibnamefont
  {Balents}},\ }\href {\doibase 10.1103/PhysRevLett.93.127202} {\bibfield
  {journal} {\bibinfo  {journal} {Phys. Rev. Lett.}\ }\textbf {\bibinfo
  {volume} {93}},\ \bibinfo {pages} {127202} (\bibinfo {year}
  {2004})}\BibitemShut {NoStop}%
\bibitem [{\citenamefont {Bombardi}\ \emph {et~al.}(2005)\citenamefont
  {Bombardi}, \citenamefont {Chapon}, \citenamefont {Margiolaki}, \citenamefont
  {Mazzoli}, \citenamefont {Gonthier}, \citenamefont {Duc},\ and\ \citenamefont
  {Radaelli}}]{Bombardi2005}%
  \BibitemOpen
  \bibfield  {author} {\bibinfo {author} {\bibfnamefont {A.}~\bibnamefont
  {Bombardi}}, \bibinfo {author} {\bibfnamefont {L.~C.}\ \bibnamefont
  {Chapon}}, \bibinfo {author} {\bibfnamefont {I.}~\bibnamefont {Margiolaki}},
  \bibinfo {author} {\bibfnamefont {C.}~\bibnamefont {Mazzoli}}, \bibinfo
  {author} {\bibfnamefont {S.}~\bibnamefont {Gonthier}}, \bibinfo {author}
  {\bibfnamefont {F.}~\bibnamefont {Duc}}, \ and\ \bibinfo {author}
  {\bibfnamefont {P.~G.}\ \bibnamefont {Radaelli}},\ }\href {\doibase
  10.1103/PhysRevB.71.220406} {\bibfield  {journal} {\bibinfo  {journal} {Phys.
  Rev. B}\ }\textbf {\bibinfo {volume} {71}},\ \bibinfo {pages} {220406}
  (\bibinfo {year} {2005})}\BibitemShut {NoStop}%
\bibitem [{\citenamefont {Nath}\ \emph {et~al.}(2008)\citenamefont {Nath},
  \citenamefont {Tsirlin}, \citenamefont {Rosner},\ and\ \citenamefont
  {Geibel}}]{Nath2008}%
  \BibitemOpen
  \bibfield  {author} {\bibinfo {author} {\bibfnamefont {R.}~\bibnamefont
  {Nath}}, \bibinfo {author} {\bibfnamefont {A.~A.}\ \bibnamefont {Tsirlin}},
  \bibinfo {author} {\bibfnamefont {H.}~\bibnamefont {Rosner}}, \ and\ \bibinfo
  {author} {\bibfnamefont {C.}~\bibnamefont {Geibel}},\ }\href {\doibase
  10.1103/PhysRevB.78.064422} {\bibfield  {journal} {\bibinfo  {journal} {Phys.
  Rev. B}\ }\textbf {\bibinfo {volume} {78}},\ \bibinfo {pages} {064422}
  (\bibinfo {year} {2008})}\BibitemShut {NoStop}%
\bibitem [{\citenamefont {Tsirlin}\ \emph {et~al.}(2008)\citenamefont
  {Tsirlin}, \citenamefont {Belik}, \citenamefont {Shpanchenko}, \citenamefont
  {Antipov}, \citenamefont {Takayama-Muromachi},\ and\ \citenamefont
  {Rosner}}]{Tsirlin2008}%
  \BibitemOpen
  \bibfield  {author} {\bibinfo {author} {\bibfnamefont {A.~A.}\ \bibnamefont
  {Tsirlin}}, \bibinfo {author} {\bibfnamefont {A.~A.}\ \bibnamefont {Belik}},
  \bibinfo {author} {\bibfnamefont {R.~V.}\ \bibnamefont {Shpanchenko}},
  \bibinfo {author} {\bibfnamefont {E.~V.}\ \bibnamefont {Antipov}}, \bibinfo
  {author} {\bibfnamefont {E.}~\bibnamefont {Takayama-Muromachi}}, \ and\
  \bibinfo {author} {\bibfnamefont {H.}~\bibnamefont {Rosner}},\ }\href
  {\doibase 10.1103/PhysRevB.77.092402} {\bibfield  {journal} {\bibinfo
  {journal} {Phys. Rev. B}\ }\textbf {\bibinfo {volume} {77}},\ \bibinfo
  {pages} {092402} (\bibinfo {year} {2008})}\BibitemShut {NoStop}%
\bibitem [{\citenamefont {Oka}\ \emph {et~al.}(2008)\citenamefont {Oka},
  \citenamefont {Yamada}, \citenamefont {Azuma}, \citenamefont {Takeshita},
  \citenamefont {Satoh}, \citenamefont {Koda}, \citenamefont {Kadono},
  \citenamefont {Takano},\ and\ \citenamefont {Shimakawa}}]{Oka2008}%
  \BibitemOpen
  \bibfield  {author} {\bibinfo {author} {\bibfnamefont {K.}~\bibnamefont
  {Oka}}, \bibinfo {author} {\bibfnamefont {I.}~\bibnamefont {Yamada}},
  \bibinfo {author} {\bibfnamefont {M.}~\bibnamefont {Azuma}}, \bibinfo
  {author} {\bibfnamefont {S.}~\bibnamefont {Takeshita}}, \bibinfo {author}
  {\bibfnamefont {K.~H.}\ \bibnamefont {Satoh}}, \bibinfo {author}
  {\bibfnamefont {A.}~\bibnamefont {Koda}}, \bibinfo {author} {\bibfnamefont
  {R.}~\bibnamefont {Kadono}}, \bibinfo {author} {\bibfnamefont
  {M.}~\bibnamefont {Takano}}, \ and\ \bibinfo {author} {\bibfnamefont
  {Y.}~\bibnamefont {Shimakawa}},\ }\href {\doibase 10.1021/ic800649a}
  {\bibfield  {journal} {\bibinfo  {journal} {Inorganic Chemistry}\ }\textbf
  {\bibinfo {volume} {47}},\ \bibinfo {pages} {7355} (\bibinfo {year}
  {2008})},\ \bibinfo {note} {pMID: 18642895}\BibitemShut {NoStop}%
\bibitem [{\citenamefont {Jiang}\ \emph {et~al.}(2012)\citenamefont {Jiang},
  \citenamefont {Yao},\ and\ \citenamefont {Balents}}]{Jiang2012}%
  \BibitemOpen
  \bibfield  {author} {\bibinfo {author} {\bibfnamefont {H.-C.}\ \bibnamefont
  {Jiang}}, \bibinfo {author} {\bibfnamefont {H.}~\bibnamefont {Yao}}, \ and\
  \bibinfo {author} {\bibfnamefont {L.}~\bibnamefont {Balents}},\ }\href
  {\doibase 10.1103/PhysRevB.86.024424} {\bibfield  {journal} {\bibinfo
  {journal} {Phys. Rev. B}\ }\textbf {\bibinfo {volume} {86}},\ \bibinfo
  {pages} {024424} (\bibinfo {year} {2012})}\BibitemShut {NoStop}%
\bibitem [{\citenamefont {Metavitsiadis}\ \emph {et~al.}(2014)\citenamefont
  {Metavitsiadis}, \citenamefont {Sellmann},\ and\ \citenamefont
  {Eggert}}]{Metavitsiadis2014}%
  \BibitemOpen
  \bibfield  {author} {\bibinfo {author} {\bibfnamefont {A.}~\bibnamefont
  {Metavitsiadis}}, \bibinfo {author} {\bibfnamefont {D.}~\bibnamefont
  {Sellmann}}, \ and\ \bibinfo {author} {\bibfnamefont {S.}~\bibnamefont
  {Eggert}},\ }\href {\doibase 10.1103/PhysRevB.89.241104} {\bibfield
  {journal} {\bibinfo  {journal} {Phys. Rev. B}\ }\textbf {\bibinfo {volume}
  {89}},\ \bibinfo {pages} {241104} (\bibinfo {year} {2014})}\BibitemShut
  {NoStop}%
\bibitem [{\citenamefont {Mermin}\ and\ \citenamefont
  {Wagner}(1966)}]{Mermin1966}%
  \BibitemOpen
  \bibfield  {author} {\bibinfo {author} {\bibfnamefont {N.~D.}\ \bibnamefont
  {Mermin}}\ and\ \bibinfo {author} {\bibfnamefont {H.}~\bibnamefont
  {Wagner}},\ }\href {\doibase 10.1103/PhysRevLett.17.1133} {\bibfield
  {journal} {\bibinfo  {journal} {Phys. Rev. Lett.}\ }\textbf {\bibinfo
  {volume} {17}},\ \bibinfo {pages} {1133} (\bibinfo {year}
  {1966})}\BibitemShut {NoStop}%
\bibitem [{\citenamefont {Scalapino}\ \emph {et~al.}(1975)\citenamefont
  {Scalapino}, \citenamefont {Imry},\ and\ \citenamefont
  {Pincus}}]{Scalapino1975}%
  \BibitemOpen
  \bibfield  {author} {\bibinfo {author} {\bibfnamefont {D.~J.}\ \bibnamefont
  {Scalapino}}, \bibinfo {author} {\bibfnamefont {Y.}~\bibnamefont {Imry}}, \
  and\ \bibinfo {author} {\bibfnamefont {P.}~\bibnamefont {Pincus}},\ }\href
  {\doibase 10.1103/PhysRevB.11.2042} {\bibfield  {journal} {\bibinfo
  {journal} {Phys. Rev. B}\ }\textbf {\bibinfo {volume} {11}},\ \bibinfo
  {pages} {2042} (\bibinfo {year} {1975})}\BibitemShut {NoStop}%
\bibitem [{\citenamefont {Schulz}(1996)}]{Schulz1996}%
  \BibitemOpen
  \bibfield  {author} {\bibinfo {author} {\bibfnamefont {H.~J.}\ \bibnamefont
  {Schulz}},\ }\href {\doibase 10.1103/PhysRevLett.77.2790} {\bibfield
  {journal} {\bibinfo  {journal} {Phys. Rev. Lett.}\ }\textbf {\bibinfo
  {volume} {77}},\ \bibinfo {pages} {2790} (\bibinfo {year}
  {1996})}\BibitemShut {NoStop}%
\bibitem [{\citenamefont {Bocquet}\ \emph {et~al.}(2001)\citenamefont
  {Bocquet}, \citenamefont {Essler}, \citenamefont {Tsvelik},\ and\
  \citenamefont {Gogolin}}]{Bocquet2001}%
  \BibitemOpen
  \bibfield  {author} {\bibinfo {author} {\bibfnamefont {M.}~\bibnamefont
  {Bocquet}}, \bibinfo {author} {\bibfnamefont {F.~H.~L.}\ \bibnamefont
  {Essler}}, \bibinfo {author} {\bibfnamefont {A.~M.}\ \bibnamefont {Tsvelik}},
  \ and\ \bibinfo {author} {\bibfnamefont {A.~O.}\ \bibnamefont {Gogolin}},\
  }\href {\doibase 10.1103/PhysRevB.64.094425} {\bibfield  {journal} {\bibinfo
  {journal} {Phys. Rev. B}\ }\textbf {\bibinfo {volume} {64}},\ \bibinfo
  {pages} {094425} (\bibinfo {year} {2001})}\BibitemShut {NoStop}%
\bibitem [{\citenamefont {Sato}\ and\ \citenamefont
  {Oshikawa}(2004)}]{Sato2004}%
  \BibitemOpen
  \bibfield  {author} {\bibinfo {author} {\bibfnamefont {M.}~\bibnamefont
  {Sato}}\ and\ \bibinfo {author} {\bibfnamefont {M.}~\bibnamefont
  {Oshikawa}},\ }\href {\doibase 10.1103/PhysRevB.69.054406} {\bibfield
  {journal} {\bibinfo  {journal} {Phys. Rev. B}\ }\textbf {\bibinfo {volume}
  {69}},\ \bibinfo {pages} {054406} (\bibinfo {year} {2004})}\BibitemShut
  {NoStop}%
\bibitem [{\citenamefont {Okunishi}\ and\ \citenamefont
  {Suzuki}(2007)}]{Okunishi2007}%
  \BibitemOpen
  \bibfield  {author} {\bibinfo {author} {\bibfnamefont {K.}~\bibnamefont
  {Okunishi}}\ and\ \bibinfo {author} {\bibfnamefont {T.}~\bibnamefont
  {Suzuki}},\ }\href {\doibase 10.1103/PhysRevB.76.224411} {\bibfield
  {journal} {\bibinfo  {journal} {Phys. Rev. B}\ }\textbf {\bibinfo {volume}
  {76}},\ \bibinfo {pages} {224411} (\bibinfo {year} {2007})}\BibitemShut
  {NoStop}%
\bibitem [{\citenamefont {Klanj\ifmmode~\check{s}\else \v{s}\fi{}ek}\ \emph
  {et~al.}(2008)\citenamefont {Klanj\ifmmode~\check{s}\else \v{s}\fi{}ek},
  \citenamefont {Mayaffre}, \citenamefont {Berthier}, \citenamefont
  {Horvati\ifmmode~\acute{c}\else \'{c}\fi{}}, \citenamefont {Chiari},
  \citenamefont {Piovesana}, \citenamefont {Bouillot}, \citenamefont {Kollath},
  \citenamefont {Orignac}, \citenamefont {Citro},\ and\ \citenamefont
  {Giamarchi}}]{Klanjsek2008}%
  \BibitemOpen
  \bibfield  {author} {\bibinfo {author} {\bibfnamefont {M.}~\bibnamefont
  {Klanj\ifmmode~\check{s}\else \v{s}\fi{}ek}}, \bibinfo {author}
  {\bibfnamefont {H.}~\bibnamefont {Mayaffre}}, \bibinfo {author}
  {\bibfnamefont {C.}~\bibnamefont {Berthier}}, \bibinfo {author}
  {\bibfnamefont {M.}~\bibnamefont {Horvati\ifmmode~\acute{c}\else
  \'{c}\fi{}}}, \bibinfo {author} {\bibfnamefont {B.}~\bibnamefont {Chiari}},
  \bibinfo {author} {\bibfnamefont {O.}~\bibnamefont {Piovesana}}, \bibinfo
  {author} {\bibfnamefont {P.}~\bibnamefont {Bouillot}}, \bibinfo {author}
  {\bibfnamefont {C.}~\bibnamefont {Kollath}}, \bibinfo {author} {\bibfnamefont
  {E.}~\bibnamefont {Orignac}}, \bibinfo {author} {\bibfnamefont
  {R.}~\bibnamefont {Citro}}, \ and\ \bibinfo {author} {\bibfnamefont
  {T.}~\bibnamefont {Giamarchi}},\ }\href {\doibase
  10.1103/PhysRevLett.101.137207} {\bibfield  {journal} {\bibinfo  {journal}
  {Phys. Rev. Lett.}\ }\textbf {\bibinfo {volume} {101}},\ \bibinfo {pages}
  {137207} (\bibinfo {year} {2008})}\BibitemShut {NoStop}%
\bibitem [{\citenamefont {Sato}\ \emph {et~al.}(2013)\citenamefont {Sato},
  \citenamefont {Hikihara},\ and\ \citenamefont {Momoi}}]{Sato2013}%
  \BibitemOpen
  \bibfield  {author} {\bibinfo {author} {\bibfnamefont {M.}~\bibnamefont
  {Sato}}, \bibinfo {author} {\bibfnamefont {T.}~\bibnamefont {Hikihara}}, \
  and\ \bibinfo {author} {\bibfnamefont {T.}~\bibnamefont {Momoi}},\ }\href
  {\doibase 10.1103/PhysRevLett.110.077206} {\bibfield  {journal} {\bibinfo
  {journal} {Phys. Rev. Lett.}\ }\textbf {\bibinfo {volume} {110}},\ \bibinfo
  {pages} {077206} (\bibinfo {year} {2013})}\BibitemShut {NoStop}%
\bibitem [{\citenamefont {Matsumoto}\ \emph {et~al.}(2001)\citenamefont
  {Matsumoto}, \citenamefont {Yasuda}, \citenamefont {Todo},\ and\
  \citenamefont {Takayama}}]{Matsumoto2001}%
  \BibitemOpen
  \bibfield  {author} {\bibinfo {author} {\bibfnamefont {M.}~\bibnamefont
  {Matsumoto}}, \bibinfo {author} {\bibfnamefont {C.}~\bibnamefont {Yasuda}},
  \bibinfo {author} {\bibfnamefont {S.}~\bibnamefont {Todo}}, \ and\ \bibinfo
  {author} {\bibfnamefont {H.}~\bibnamefont {Takayama}},\ }\href {\doibase
  10.1103/PhysRevB.65.014407} {\bibfield  {journal} {\bibinfo  {journal} {Phys.
  Rev. B}\ }\textbf {\bibinfo {volume} {65}},\ \bibinfo {pages} {014407}
  (\bibinfo {year} {2001})}\BibitemShut {NoStop}%
\bibitem [{\citenamefont {Yasuda}\ \emph {et~al.}(2005)\citenamefont {Yasuda},
  \citenamefont {Todo}, \citenamefont {Hukushima}, \citenamefont {Alet},
  \citenamefont {Keller}, \citenamefont {Troyer},\ and\ \citenamefont
  {Takayama}}]{Yasuda2005}%
  \BibitemOpen
  \bibfield  {author} {\bibinfo {author} {\bibfnamefont {C.}~\bibnamefont
  {Yasuda}}, \bibinfo {author} {\bibfnamefont {S.}~\bibnamefont {Todo}},
  \bibinfo {author} {\bibfnamefont {K.}~\bibnamefont {Hukushima}}, \bibinfo
  {author} {\bibfnamefont {F.}~\bibnamefont {Alet}}, \bibinfo {author}
  {\bibfnamefont {M.}~\bibnamefont {Keller}}, \bibinfo {author} {\bibfnamefont
  {M.}~\bibnamefont {Troyer}}, \ and\ \bibinfo {author} {\bibfnamefont
  {H.}~\bibnamefont {Takayama}},\ }\href {\doibase
  10.1103/PhysRevLett.94.217201} {\bibfield  {journal} {\bibinfo  {journal}
  {Phys. Rev. Lett.}\ }\textbf {\bibinfo {volume} {94}},\ \bibinfo {pages}
  {217201} (\bibinfo {year} {2005})}\BibitemShut {NoStop}%
\bibitem [{\citenamefont {Sato}\ and\ \citenamefont
  {Oshikawa}(2007)}]{Sato2007}%
  \BibitemOpen
  \bibfield  {author} {\bibinfo {author} {\bibfnamefont {M.}~\bibnamefont
  {Sato}}\ and\ \bibinfo {author} {\bibfnamefont {M.}~\bibnamefont
  {Oshikawa}},\ }\href {\doibase 10.1103/PhysRevB.75.014404} {\bibfield
  {journal} {\bibinfo  {journal} {Phys. Rev. B}\ }\textbf {\bibinfo {volume}
  {75}},\ \bibinfo {pages} {014404} (\bibinfo {year} {2007})}\BibitemShut
  {NoStop}%
\bibitem [{\citenamefont {Schlappa}\ \emph {et~al.}(2012)\citenamefont
  {Schlappa}, \citenamefont {Wohlfeld}, \citenamefont {Zhou}, \citenamefont
  {Mourigal}, \citenamefont {Haverkort}, \citenamefont {Strocov}, \citenamefont
  {Hozoi}, \citenamefont {Monney}, \citenamefont {Nishimoto}, \citenamefont
  {Singh}, \citenamefont {Revcolevschi}, \citenamefont {Caux}, \citenamefont
  {Patthey}, \citenamefont {R{\o}nnow}, \citenamefont {van~den Brink},\ and\
  \citenamefont {Schmitt}}]{Schlappa2012}%
  \BibitemOpen
  \bibfield  {author} {\bibinfo {author} {\bibfnamefont {J.}~\bibnamefont
  {Schlappa}}, \bibinfo {author} {\bibfnamefont {K.}~\bibnamefont {Wohlfeld}},
  \bibinfo {author} {\bibfnamefont {K.~J.}\ \bibnamefont {Zhou}}, \bibinfo
  {author} {\bibfnamefont {M.}~\bibnamefont {Mourigal}}, \bibinfo {author}
  {\bibfnamefont {M.~W.}\ \bibnamefont {Haverkort}}, \bibinfo {author}
  {\bibfnamefont {V.~N.}\ \bibnamefont {Strocov}}, \bibinfo {author}
  {\bibfnamefont {L.}~\bibnamefont {Hozoi}}, \bibinfo {author} {\bibfnamefont
  {C.}~\bibnamefont {Monney}}, \bibinfo {author} {\bibfnamefont
  {S.}~\bibnamefont {Nishimoto}}, \bibinfo {author} {\bibfnamefont
  {S.}~\bibnamefont {Singh}}, \bibinfo {author} {\bibfnamefont
  {A.}~\bibnamefont {Revcolevschi}}, \bibinfo {author} {\bibfnamefont {J.-S.}\
  \bibnamefont {Caux}}, \bibinfo {author} {\bibfnamefont {L.}~\bibnamefont
  {Patthey}}, \bibinfo {author} {\bibfnamefont {H.~M.}\ \bibnamefont
  {R{\o}nnow}}, \bibinfo {author} {\bibfnamefont {J.}~\bibnamefont {van~den
  Brink}}, \ and\ \bibinfo {author} {\bibfnamefont {T.}~\bibnamefont
  {Schmitt}},\ }\href {http://dx.doi.org/10.1038/nature10974} {\bibfield
  {journal} {\bibinfo  {journal} {Nature}\ }\textbf {\bibinfo {volume} {485}},\
  \bibinfo {pages} {82 EP } (\bibinfo {year} {2012})}\BibitemShut {NoStop}%
\bibitem [{\citenamefont {Kohno}\ \emph {et~al.}(2007)\citenamefont {Kohno},
  \citenamefont {Starykh},\ and\ \citenamefont {Balents}}]{Kohno2007}%
  \BibitemOpen
  \bibfield  {author} {\bibinfo {author} {\bibfnamefont {M.}~\bibnamefont
  {Kohno}}, \bibinfo {author} {\bibfnamefont {O.~A.}\ \bibnamefont {Starykh}},
  \ and\ \bibinfo {author} {\bibfnamefont {L.}~\bibnamefont {Balents}},\ }\href
  {http://dx.doi.org/10.1038/nphys749} {\bibfield  {journal} {\bibinfo
  {journal} {Nature Physics}\ }\textbf {\bibinfo {volume} {3}},\ \bibinfo
  {pages} {790 EP } (\bibinfo {year} {2007})},\ \bibinfo {note}
  {article}\BibitemShut {NoStop}%
\bibitem [{\citenamefont {Coldea}\ \emph {et~al.}(2002)\citenamefont {Coldea},
  \citenamefont {Tennant}, \citenamefont {Habicht}, \citenamefont {Smeibidl},
  \citenamefont {Wolters},\ and\ \citenamefont {Tylczynski}}]{Coldea2002}%
  \BibitemOpen
  \bibfield  {author} {\bibinfo {author} {\bibfnamefont {R.}~\bibnamefont
  {Coldea}}, \bibinfo {author} {\bibfnamefont {D.~A.}\ \bibnamefont {Tennant}},
  \bibinfo {author} {\bibfnamefont {K.}~\bibnamefont {Habicht}}, \bibinfo
  {author} {\bibfnamefont {P.}~\bibnamefont {Smeibidl}}, \bibinfo {author}
  {\bibfnamefont {C.}~\bibnamefont {Wolters}}, \ and\ \bibinfo {author}
  {\bibfnamefont {Z.}~\bibnamefont {Tylczynski}},\ }\href {\doibase
  10.1103/PhysRevLett.88.137203} {\bibfield  {journal} {\bibinfo  {journal}
  {Phys. Rev. Lett.}\ }\textbf {\bibinfo {volume} {88}},\ \bibinfo {pages}
  {137203} (\bibinfo {year} {2002})}\BibitemShut {NoStop}%
\bibitem [{\citenamefont {Mourigal}\ \emph {et~al.}(2013)\citenamefont
  {Mourigal}, \citenamefont {Enderle}, \citenamefont {Kl{\"o}pperpieper},
  \citenamefont {Caux}, \citenamefont {Stunault},\ and\ \citenamefont
  {R{\o}nnow}}]{Mourigal2013}%
  \BibitemOpen
  \bibfield  {author} {\bibinfo {author} {\bibfnamefont {M.}~\bibnamefont
  {Mourigal}}, \bibinfo {author} {\bibfnamefont {M.}~\bibnamefont {Enderle}},
  \bibinfo {author} {\bibfnamefont {A.}~\bibnamefont {Kl{\"o}pperpieper}},
  \bibinfo {author} {\bibfnamefont {J.-S.}\ \bibnamefont {Caux}}, \bibinfo
  {author} {\bibfnamefont {A.}~\bibnamefont {Stunault}}, \ and\ \bibinfo
  {author} {\bibfnamefont {H.~M.}\ \bibnamefont {R{\o}nnow}},\ }\href
  {http://dx.doi.org/10.1038/nphys2652} {\bibfield  {journal} {\bibinfo
  {journal} {Nature Physics}\ }\textbf {\bibinfo {volume} {9}},\ \bibinfo
  {pages} {435 EP } (\bibinfo {year} {2013})},\ \bibinfo {note}
  {article}\BibitemShut {NoStop}%
\bibitem [{\citenamefont {Epstein}\ \emph {et~al.}(1971)\citenamefont
  {Epstein}, \citenamefont {Etemad}, \citenamefont {Garito},\ and\
  \citenamefont {Heeger}}]{Epstein1971}%
  \BibitemOpen
  \bibfield  {author} {\bibinfo {author} {\bibfnamefont {A.}~\bibnamefont
  {Epstein}}, \bibinfo {author} {\bibfnamefont {S.}~\bibnamefont {Etemad}},
  \bibinfo {author} {\bibfnamefont {A.}~\bibnamefont {Garito}}, \ and\ \bibinfo
  {author} {\bibfnamefont {A.}~\bibnamefont {Heeger}},\ }\href {\doibase
  https://doi.org/10.1016/0038-1098(71)90094-9} {\bibfield  {journal} {\bibinfo
   {journal} {Solid State Communications}\ }\textbf {\bibinfo {volume} {9}},\
  \bibinfo {pages} {1803 } (\bibinfo {year} {1971})}\BibitemShut {NoStop}%
\bibitem [{\citenamefont {Renard}\ \emph {et~al.}(1987)\citenamefont {Renard},
  \citenamefont {Verdaguer}, \citenamefont {Regnault}, \citenamefont
  {Erkelens}, \citenamefont {Rossat-Mignod},\ and\ \citenamefont
  {Stirling}}]{Renard1987}%
  \BibitemOpen
  \bibfield  {author} {\bibinfo {author} {\bibfnamefont {J.~P.}\ \bibnamefont
  {Renard}}, \bibinfo {author} {\bibfnamefont {M.}~\bibnamefont {Verdaguer}},
  \bibinfo {author} {\bibfnamefont {L.~P.}\ \bibnamefont {Regnault}}, \bibinfo
  {author} {\bibfnamefont {W.~A.~C.}\ \bibnamefont {Erkelens}}, \bibinfo
  {author} {\bibfnamefont {J.}~\bibnamefont {Rossat-Mignod}}, \ and\ \bibinfo
  {author} {\bibfnamefont {W.~G.}\ \bibnamefont {Stirling}},\ }\href
  {http://stacks.iop.org/0295-5075/3/i=8/a=013} {\bibfield  {journal} {\bibinfo
   {journal} {EPL (Europhysics Letters)}\ }\textbf {\bibinfo {volume} {3}},\
  \bibinfo {pages} {945} (\bibinfo {year} {1987})}\BibitemShut {NoStop}%
\bibitem [{\citenamefont {Katsumata}\ \emph {et~al.}(1989)\citenamefont
  {Katsumata}, \citenamefont {Hori}, \citenamefont {Takeuchi}, \citenamefont
  {Date}, \citenamefont {Yamagishi},\ and\ \citenamefont
  {Renard}}]{Katsumata1989}%
  \BibitemOpen
  \bibfield  {author} {\bibinfo {author} {\bibfnamefont {K.}~\bibnamefont
  {Katsumata}}, \bibinfo {author} {\bibfnamefont {H.}~\bibnamefont {Hori}},
  \bibinfo {author} {\bibfnamefont {T.}~\bibnamefont {Takeuchi}}, \bibinfo
  {author} {\bibfnamefont {M.}~\bibnamefont {Date}}, \bibinfo {author}
  {\bibfnamefont {A.}~\bibnamefont {Yamagishi}}, \ and\ \bibinfo {author}
  {\bibfnamefont {J.~P.}\ \bibnamefont {Renard}},\ }\href {\doibase
  10.1103/PhysRevLett.63.86} {\bibfield  {journal} {\bibinfo  {journal} {Phys.
  Rev. Lett.}\ }\textbf {\bibinfo {volume} {63}},\ \bibinfo {pages} {86}
  (\bibinfo {year} {1989})}\BibitemShut {NoStop}%
\bibitem [{\citenamefont {Gadet}\ \emph {et~al.}(1991)\citenamefont {Gadet},
  \citenamefont {Verdaguer}, \citenamefont {Briois}, \citenamefont {Gleizes},
  \citenamefont {Renard}, \citenamefont {Beauvillain}, \citenamefont
  {Chappert}, \citenamefont {Goto}, \citenamefont {Le~Dang},\ and\
  \citenamefont {Veillet}}]{Gadet1991}%
  \BibitemOpen
  \bibfield  {author} {\bibinfo {author} {\bibfnamefont {V.}~\bibnamefont
  {Gadet}}, \bibinfo {author} {\bibfnamefont {M.}~\bibnamefont {Verdaguer}},
  \bibinfo {author} {\bibfnamefont {V.}~\bibnamefont {Briois}}, \bibinfo
  {author} {\bibfnamefont {A.}~\bibnamefont {Gleizes}}, \bibinfo {author}
  {\bibfnamefont {J.~P.}\ \bibnamefont {Renard}}, \bibinfo {author}
  {\bibfnamefont {P.}~\bibnamefont {Beauvillain}}, \bibinfo {author}
  {\bibfnamefont {C.}~\bibnamefont {Chappert}}, \bibinfo {author}
  {\bibfnamefont {T.}~\bibnamefont {Goto}}, \bibinfo {author} {\bibfnamefont
  {K.}~\bibnamefont {Le~Dang}}, \ and\ \bibinfo {author} {\bibfnamefont
  {P.}~\bibnamefont {Veillet}},\ }\href {\doibase 10.1103/PhysRevB.44.705}
  {\bibfield  {journal} {\bibinfo  {journal} {Phys. Rev. B}\ }\textbf {\bibinfo
  {volume} {44}},\ \bibinfo {pages} {705} (\bibinfo {year} {1991})}\BibitemShut
  {NoStop}%
\bibitem [{\citenamefont {Darriet}\ and\ \citenamefont
  {Regnault}(1993)}]{Darriet1993}%
  \BibitemOpen
  \bibfield  {author} {\bibinfo {author} {\bibfnamefont {J.}~\bibnamefont
  {Darriet}}\ and\ \bibinfo {author} {\bibfnamefont {L.}~\bibnamefont
  {Regnault}},\ }\href {\doibase https://doi.org/10.1016/0038-1098(93)90455-V}
  {\bibfield  {journal} {\bibinfo  {journal} {Solid State Communications}\
  }\textbf {\bibinfo {volume} {86}},\ \bibinfo {pages} {409 } (\bibinfo {year}
  {1993})}\BibitemShut {NoStop}%
\bibitem [{\citenamefont {Sakaguchi}\ \emph {et~al.}(1996)\citenamefont
  {Sakaguchi}, \citenamefont {Kakurai}, \citenamefont {Yokoo},\ and\
  \citenamefont {Akimitsu}}]{Sakaguchi1996}%
  \BibitemOpen
  \bibfield  {author} {\bibinfo {author} {\bibfnamefont {T.}~\bibnamefont
  {Sakaguchi}}, \bibinfo {author} {\bibfnamefont {K.}~\bibnamefont {Kakurai}},
  \bibinfo {author} {\bibfnamefont {T.}~\bibnamefont {Yokoo}}, \ and\ \bibinfo
  {author} {\bibfnamefont {J.}~\bibnamefont {Akimitsu}},\ }\href {\doibase
  10.1143/JPSJ.65.3025} {\bibfield  {journal} {\bibinfo  {journal} {Journal of
  the Physical Society of Japan}\ }\textbf {\bibinfo {volume} {65}},\ \bibinfo
  {pages} {3025} (\bibinfo {year} {1996})}\BibitemShut {NoStop}%
\bibitem [{\citenamefont {Lukyanov}\ and\ \citenamefont
  {Zamolodchikov}(1997)}]{Lukyanov1997}%
  \BibitemOpen
  \bibfield  {author} {\bibinfo {author} {\bibfnamefont {S.}~\bibnamefont
  {Lukyanov}}\ and\ \bibinfo {author} {\bibfnamefont {A.}~\bibnamefont
  {Zamolodchikov}},\ }\href {\doibase
  https://doi.org/10.1016/S0550-3213(97)00123-5} {\bibfield  {journal}
  {\bibinfo  {journal} {Nuclear Physics B}\ }\textbf {\bibinfo {volume}
  {493}},\ \bibinfo {pages} {571 } (\bibinfo {year} {1997})}\BibitemShut
  {NoStop}%
\bibitem [{\citenamefont {Lukyanov}(1999)}]{Lukyanov1999}%
  \BibitemOpen
  \bibfield  {author} {\bibinfo {author} {\bibfnamefont {S.}~\bibnamefont
  {Lukyanov}},\ }\href {\doibase 10.1103/PhysRevB.59.11163} {\bibfield
  {journal} {\bibinfo  {journal} {Phys. Rev. B}\ }\textbf {\bibinfo {volume}
  {59}},\ \bibinfo {pages} {11163} (\bibinfo {year} {1999})}\BibitemShut
  {NoStop}%
\bibitem [{\citenamefont {Lukyanov}\ and\ \citenamefont
  {Terras}(2003)}]{Lukyanov2003}%
  \BibitemOpen
  \bibfield  {author} {\bibinfo {author} {\bibfnamefont {S.}~\bibnamefont
  {Lukyanov}}\ and\ \bibinfo {author} {\bibfnamefont {V.}~\bibnamefont
  {Terras}},\ }\href {\doibase https://doi.org/10.1016/S0550-3213(02)01141-0}
  {\bibfield  {journal} {\bibinfo  {journal} {Nuclear Physics B}\ }\textbf
  {\bibinfo {volume} {654}},\ \bibinfo {pages} {323 } (\bibinfo {year}
  {2003})}\BibitemShut {NoStop}%
\bibitem [{\citenamefont {Hikihara}\ and\ \citenamefont
  {Furusaki}(1998)}]{Hikihara1998}%
  \BibitemOpen
  \bibfield  {author} {\bibinfo {author} {\bibfnamefont {T.}~\bibnamefont
  {Hikihara}}\ and\ \bibinfo {author} {\bibfnamefont {A.}~\bibnamefont
  {Furusaki}},\ }\href {\doibase 10.1103/PhysRevB.58.R583} {\bibfield
  {journal} {\bibinfo  {journal} {Phys. Rev. B}\ }\textbf {\bibinfo {volume}
  {58}},\ \bibinfo {pages} {R583} (\bibinfo {year} {1998})}\BibitemShut
  {NoStop}%
\bibitem [{\citenamefont {Hikihara}\ and\ \citenamefont
  {Furusaki}(2004)}]{Hikihara2004}%
  \BibitemOpen
  \bibfield  {author} {\bibinfo {author} {\bibfnamefont {T.}~\bibnamefont
  {Hikihara}}\ and\ \bibinfo {author} {\bibfnamefont {A.}~\bibnamefont
  {Furusaki}},\ }\href {\doibase 10.1103/PhysRevB.69.064427} {\bibfield
  {journal} {\bibinfo  {journal} {Phys. Rev. B}\ }\textbf {\bibinfo {volume}
  {69}},\ \bibinfo {pages} {064427} (\bibinfo {year} {2004})}\BibitemShut
  {NoStop}%
\bibitem [{\citenamefont {Takayoshi}\ and\ \citenamefont
  {Sato}(2010)}]{Takayoshi2010}%
  \BibitemOpen
  \bibfield  {author} {\bibinfo {author} {\bibfnamefont {S.}~\bibnamefont
  {Takayoshi}}\ and\ \bibinfo {author} {\bibfnamefont {M.}~\bibnamefont
  {Sato}},\ }\href {\doibase 10.1103/PhysRevB.82.214420} {\bibfield  {journal}
  {\bibinfo  {journal} {Phys. Rev. B}\ }\textbf {\bibinfo {volume} {82}},\
  \bibinfo {pages} {214420} (\bibinfo {year} {2010})}\BibitemShut {NoStop}%
\end{thebibliography}
\end{document}